\documentclass[aps,pre,preprint,superscriptaddress, nofootinbib, longbibliography, showkeys]{revtex4-1}
\usepackage{amsmath} 
\usepackage{graphicx}
\usepackage{hyperref}
\usepackage{braket}

\newcommand{\be}{\begin{equation}} \newcommand{\ee}{\end{equation}}

\newcommand{\ben}{\begin{enumerate}} \newcommand{\een}{\end{enumerate}}
\newcommand{\bc}{\begin{center}} \newcommand{\ec}{\end{center}}
\newcommand{\bi}{\begin{itemize}} \newcommand{\ei}{\end{itemize}}

\DeclareMathOperator{\argmin}{arg\,min}
\usepackage[bitstream-charter, greekfamily=default]{mathdesign}

\begin{document}

\title{May's Instability in Large Economies}
\author{Jos\'{e} Moran}
\affiliation{Centre d'Analyse et de Math\'{e}matique Sociales, EHESS, 54 Boulevard Raspail, 75006 Paris}
\author{Jean-Philippe Bouchaud}
\affiliation{Capital Fund Management, 23 Rue de l'Universit\'{e}, 75007 Paris}
\begin{abstract}
Will a large economy be stable? Building on Robert May's original argument for large ecosystems, we conjecture that evolutionary and behavioural forces conspire to drive the economy towards marginal stability. We study networks of firms in which inputs for production are not easily substitutable, as in several real-world supply chains. Relying on results from Random Matrix Theory, we argue that such networks generically become dysfunctional when their size increases, when the heterogeneity between firms becomes too strong or when substitutability of their production inputs is reduced. At marginal stability and for large heterogeneities, we find that the distribution of firm sizes develops a power-law tail, as observed empirically. Crises can be triggered by small idiosyncratic shocks, which lead to ``avalanches'' of defaults characterized by a power-law distribution of total output losses.  This scenario would naturally explain the well-known ``small shocks, large business cycles'' puzzle, as anticipated long ago by Bak, Chen, Scheinkman and Woodford.
\end{abstract}
\maketitle

\section{Introduction}

Why is the output of large economies so volatile? Why do small idiosyncratic fluctuations lead to large business cycles? These questions have been at the forefront of economic research for decades~\cite{long1983real,cochrane1994shocks,bernanke1994financial,Lucas}. 

Naively, one would expect that the fluctuations of an economy made of $N$ independent sectors should decay rather quickly, as $N^{-1/2}$~\cite{Lucas,dupor1999aggregation} because of the Central Limit Theorem. In order to explain why fluctuations survive at the aggregate level, three families of explanations have been proposed in the literature. The first one is that aggregate fluctuations are driven by global shocks, that affect all firms/sectors simultaneously. However, it is often not clear what these shocks might be\footnote{As Cochrane quipped~\cite{cochrane1994shocks} {\it What shocks are responsible for economic fluctuations? Despite at least two hundred years in which economists have observed fluctuations in economic activity, we still are not sure.}} and, when identified, they appear too small to be responsible for the observed volatility of the aggregate industrial production. Bernanke et al.~\cite{bernanke1994financial} have called this the {\it small shocks, large cycles puzzle}. One interesting possibility is that these shocks are self-fulfilling prophecies~\cite{farmer1999macroeconomics}, perhaps due to collective opinion shifts or trust collapse, see e.g.~\cite{Brock_2001,Bouchaud_2013,Anand_2013,da_Gama_Batista_2015} for various strands of literature on the subject. 

Another resolution has been proposed by Gabaix~\cite{gabaix2011granular} and, in a slightly different context, by Wyart \& Bouchaud~\cite{wyart2003statistical}, see also~\cite{sutton2002variance}. The argument is that the fat-tailed distribution of sizes of independent firms/sectors slows down the regression of fluctuations from the standard $N^{-1/2}$ behaviour to $N^{-\alpha}$, with $\alpha \leq 1/2$ related to the tail exponent of the distribution. Although some empirical support for this scenario has been put forth~\cite{gabaix2011granular,di_Giovanni_2017}, other works suggest that {\it network effects} are in fact dominant~\cite{acemoglu2012network,acemoglu2015networks,acemoglu2016networks}, as idiosyncratic shocks can cascade along the input-output network and eventually become macroscopic\footnote{Similar network effects have been argued to be at the origin of system-wide breakdowns of the banking sector.~\cite{HaldaneMay2011,Gai2010,Tasca_2011,Caccioli2017}.}. One particular stigma of these network effects is the strong co-variation of fluctuations across different sectors~\cite{Foerster_2008} -- but see also~\cite{Carvalho_2010}.

While the cascade story is enticing, the baseline Cobb-Douglas network model proposed by Acemoglu, Carvalho et al.~\cite{acemoglu2012network,long1983real} is, in our view, not convincing. Indeed, the only way to escape the $N^{-1/2}$ decay of fluctuations within this framework is to assume that the supply network is very unbalanced, i.e. that a few sectors are crucial suppliers to the whole economy~\cite{acemoglu2013network}. This somehow throws the baby with the bathwater, as it re-introduces the idea of aggregate shocks in disguise. A possible way out was proposed in Ref.~\cite{bonart2014instabilities}: by introducing myopia and frictions in the Cobb-Douglas network model, it was found that the general equilibrium solution of Acemoglu et al.~\cite{acemoglu2012network} is only stable in a certain region of parameters, outside of which large fluctuations emerge {\it endogenously}, i.e. without any microscopic shocks. These fluctuations arise from a breakdown of coordination between the different sectors and illustrate how some mechanisms present in the real world may cause the economy to be intrinsically unstable.

This idea (that the economy may in itself be unstable and turbulent) was in fact already mentioned in 1948 by Hawkins~\cite{hawkins1948some} (see also~\cite{hawkins1949note}) and picked up again by Bak, Chen, Scheinkman and Woodford~\cite{bak1993aggregate,scheinkman1994self} in the context of ``self-organized critical'' (SOC) states in complex systems~\cite{bak2013nature}. In such a state, small microscopic perturbations give rise to macroscopic fluctuations --- like avalanches in sand piles. 

Similar ideas have emerged in the context of theoretical ecology. In his seminal paper, {\it Will a Large Complex System be Stable?}, May~\cite{may1972will} argued that a large number of very different species can lead an ecosystem to instabilities and mass extinctions. May's paradigm has recently been made much more explicit in the context of a generalized Lotka-Volterra model in Refs.~\cite{Bunin_2017,bunin2016interaction,biroli2018marginally}, where it is shown that the system indeed spontaneously evolves towards a marginally stable state that is anomalously sensitive to small perturbations. Unfortunately, this stream of ideas has not gained much traction in the economics literature, perhaps for lack of a convincing modelling framework. (See however~\cite{HaldaneMay2011,Gai2010,Tasca_2011,Caccioli2017} for financial network models with explicit references to ecosystems, and~\cite{Nirei_2019} for a very recent contribution.)

The aim of the present work is to present an economically motivated model where marginal stability appears naturally, and leads to an amplification of small, idiosyncratic shocks along the input-output network. Interestingly, our model is closely related to the ecological models alluded above, and builds upon the classic --- but still extremely active --- field of Random Matrix Theory, that describes the statistical properties of the eigenvalues/eigenvectors of certain families of random matrices, here related to the input-output matrix. In particular, the \textit{feasability} of an equilibrium, defined as the existence of an economically sound set of prices and production outputs, depends strongly on the eigenvalues of such a matrix. We define here the stability of the economy as the resilience of such an equilibrium to idiosyncratic shocks. We find that for a fixed number of firms $N$, increased interlinkages, profit maximisation and/or reduced substitutability drive the system at the edge of instability. Similarly, increasing the number of firms at fixed productivity also leads to a critical state. At criticality, small idiosyncratic shocks can lead to bankruptcy avalanches, which, depending on the topology of the network and the heterogeneity of firms' productivity, can be either small and localized or system-wide, with all possible gradations.

In the present paper, we only describe the equilibrium (or absence thereof) aspects of our model, leaving the analysis of its truly dynamical features --- crucial when crises occur --- for a forthcoming publication~\cite{ustocome}. 

\section{The Model}

We consider $N$ firms with a given input-output network determined by the technology available to firms. The ``technology network'' is a directed graph where nodes represent firms and where a directed edge $j\to i$ exists if $i$ needs goods produced by $j$ for its own production. Note that this framework allows for self-loops, with an edge $i\to i$ existing if a firm produces one of its own inputs. The node $i=0$ conventionally represents households, and supplies firms with labour while consuming a part of their output. The link $j \to i$ carries a ``stoichiometric weight'' $J_{ij}$, measuring the number of $j$ goods needed to make an unit of $i$'s production. The set of suppliers of $i$ is thus given by $\{j/ J_{ij}\neq 0\}$, while the set of clients is $\{j/ J_{ji}\neq 0\}$. The production of $i$, $\pi_i$, is given by a so-called CES (Constant Elasticity of Substitution) function, which reads~\cite{Raval_2019}\footnote{For a more in-depth exploration of these production functions, see Appendix~\ref{sub:appendix}.}
\begin{equation}\label{eq:def_ces}
\pi_i = z_i \left(\sum_j a_{ij} \left(\frac{J_{ij}}{Q_{ij}}\right)^{\frac{1}{q}}\right)^{-q}, \quad \mbox{with} \quad \sum_{j} a_{ij}=1,
\end{equation}
where $z_i$ is the firm's productivity, $Q_{ij}$ is the number of goods firm $i$ buys from firm $j$ and $a_{ij} \geq 0$ are weight parameters. The parameter $q$ measures the global substitutability of the different inputs. When $q \to 0$, no substitutes are available and Eq. (\ref{eq:def_ces}) reduces to the classical Leontief production function:
\begin{equation}\label{eq:def_leontieff}
    \pi_i = z_i \min_{j} \left(\frac{Q_{ij}}{J_{ij}}\right)
\end{equation}
The Cobb-Douglas function $\pi_i= z_i \prod_j (Q_{ij}/J_{ij})^{a_{ij}}$ corresponds to $q \to \infty$ and is often used to describe the average aggregate production of economic sectors~\cite{long1983real}, or of the economy as a whole. In a Cobb-Douglas economy, the loss of a fraction $f$ of good $j$ can always be compensated by an increase of any other good $k$ by a factor $1/f^{a_{ij}/a_{ik}}$. In the Leontief case, on the other hand, the loss of a fraction $f$ of good $j$ cannot be compensated and translates to an immediate loss of the same fraction $f$ of total production $\pi_i$. It models a situation where redundancy is costly. Firms therefore choose their suppliers with parsimony and cannot ``rewire'' (i.e. find alternative suppliers) on short time scales in the real economy. For example, in the aftermath of the 2011 tsunami and Fukushima Daiichi nuclear power plant disaster, the shortage of a few, seemingly unimportant components had a severe impact on the car industry~\cite{reed_simon_2011,reuters_japan_2011}, very dependent on products manufactured in Japan. The incident highlighted how competition led firms to have a very tight supply-chain strategy, as in the words of an observer: ``In the race to provide better quality at lower prices, manufacturers picked very narrow, optimized supply chains'' which caused them to be very dependent on the ``one supplier that had the best product at the lowest price''~\cite{fisher_hbs_2011}. At the time of the disruption, firms had to swiftly re-think their supply-chain strategy and large-scale rewirings of the production network took place. The influence of possible rewirings is beyond the scope of this paper, but for preliminary work in that direction see~\cite{colon2019}\footnote{One could consider a case where firms have several possible suppliers $j$ within the same sector $J$ and write $\pi_i = z_i \min_{J} \max_{j \in J} \left(\frac{Q_{ij}}{J_{ij}}\right)$. This extension will be studied at a later stage of the project.}. In the following, we will for simplicity focus on the extreme case of a Leontief production function $q \to 0$, but will show that our results hold true for in a range $q \in [0,q_c]$, where the critical value $q_c$ depends on the network and on the productivities.    

Calling $p_i$ the price of the goods produced by firm $i$, its profit $\mathcal{P}_i$ reads:
\begin{equation}\label{eq:profit}
\mathcal{P}_i = p_i\pi_i - \sum_{j\neq 0  }Q_{ij}p_j-Q_{i0}p_0,
\end{equation}
where $p_0$ is the labour wage. Optimizing the profit with respect to all inputs $Q_{ij}$ (including labour $Q_{i0}$) leads, for $q \to 0$, to the condition:
\begin{equation}\label{eq:output_def}
\forall (i,j),  \exists \gamma_i \geq 0 \quad \mbox{ s.t. } \quad Q_{ij} = \gamma_i J_{ij} 
\end{equation}
which can also be interpreted as saying that given an output level $\gamma_i := \pi_i/z_i$, the optimal choice for inputs $Q_{ij}$ is to pick them proportionally to their stoichiometric weight, as buying more would result in waste. In this case, profit can be written as
\begin{equation}\label{eq:new_profit}
\mathcal{P}_i = \gamma_i \left(z_i p_i -\sum_{j\neq 0}J_{ij}p_j - J_{i0}p_0\right).
\end{equation}

We now assume that households' optimal consumption of good $i$, given a certain utility function and a vector of prices, are given by $C_i > 0$\footnote{For example, for a utility function 
$\mathcal{U}=\sum_i \theta_i \log(C_i)$ and a certain budget $B$, the optimal consumption $C_i$ is:
\begin{equation}\label{eq:consumption}
C_i = \frac{B}{\sum_j \theta_j} \frac{\theta_i}{p_i}:= \frac{\Gamma_i}{p_i}
\end{equation}
but any other type of utility function would work in our model, as long as consumption levels be strictly positive.}.

As standard in the literature, we now assume:
\begin{itemize} 
\item Market Clearing, i.e. every good that is produced is either consumed by households or bought by other firms for their own production. Hence:
\begin{equation}\label{eq:excess_prod}
\pi_i = \sum_{j\neq 0} Q_{ji} + C_i \longrightarrow z_i\gamma_i - \sum_{j\neq 0} J_{ji}\gamma_j = C_i \quad (> 0).
\end{equation}
\item Competitive Equilibrium, i.e. competition drives profits to zero. Hence:
\begin{equation}\label{eq:excess_prof}
\mathcal{P}_i = 0 \longrightarrow  z_i p_i - \sum_{j\neq 0}J_{ij}p_j = V_i \quad (> 0),
\end{equation}
where we have defined $V_i=J_{i0}p_0$ and imposed that $\gamma_i \neq 0$, $\forall i$ (otherwise Eq. (\ref{eq:excess_prod}) cannot be satisfied). One could also model firms attempting to impose mark-ups to reach a positive profit equal to a fraction $\varphi_i$ of its sales $z_i \gamma_i p_i$. This simply amounts to shifting $z_i$ to $z_i(1-\varphi_i)$ in Eq. (\ref{eq:excess_prof}).
\end{itemize}

Now, in order for the equilibrium to make sense, the solutions to Eqs.  (\ref{eq:excess_prod},\ref{eq:excess_prof}) must be such that $\gamma_i > 0$ and $p_i > 0$, $\forall i$; i.e. that equilibrium prices and quantities must be strictly positive. As first noted by Hawkins \& Simon~\cite{hawkins1949note}, this is not automatic and requires the matrix $\mathbf{M}$, defined by $(\mathbf{M})_{ij}=z_i\delta_{ij}-J_{ij}$ to be a so-called ``M-matrix''\footnote{Note that if $\mathbf{M}$ is an M-matrix, $\mathbf{M}^t$ is also an M-matrix. An interesting property of an M-matrix is that all the elements of its inverse are non negative.}, i.e. such that {\it all its eigenvalues have non-negative real parts}~\cite{fiedler1962matrices}. Therefore some conditions on productivities and linkages must be fulfilled for the economy to work.

This condition is the equivalent, in an ecological context, of May's stability criterion that allows the equilibrium population of all species to be strictly positive~\cite{Bunin_2017,bunin2016interaction}. Rather interestingly, Eqs.  (\ref{eq:excess_prod},\ref{eq:excess_prof}) are identical, {\it mutatis mutandis}, to the equation determining the equilibrium size of species in a generalized Lotka-Volterra model~\cite{biroli2018marginally}. 

In the case of a more general CES production function with $q \geq 0$, the competitive equilibrium equation reads:
\begin{equation}\label{eq:excess_prof2}
\left(z_i p_i\right)^{\zeta} - \sum_{j\neq 0} a_{ij}^{q\zeta} \left(J_{ij} p_j\right)^{\zeta} = V_i \quad (> 0), \qquad \zeta:=\frac{1}{1+q}
\end{equation}
which boils down to Eq. (\ref{eq:excess_prof}) when $q=0$ (for a detailed proof, see Eq. \eqref{eq:competitive_eq_q} in Appendix \ref{sub:appendixB}). Interestingly, setting $\widehat p_i = p_i^{\zeta}$, one finds again that the condition for an admissible equilibrium is that the matrix $(\widehat{\mathbf{M}})_{ij}=z_i^\zeta \delta_{ij}- a_{ij}^{q \zeta} J_{ij}^\zeta$ is an M-matrix. 

Note that since $\sum_{j \neq 0} a_{ij} < 1$, the Perron-Frobenius theorem ensures that Cobb-Douglas networked economies (such as those considered in Acemoglu, Carvalho et al.~\cite{acemoglu2012network} and corresponding to $q \to \infty$), {\it always} have an admissible equilibrium, for any network and any productivities. Therefore, the type of shock propagation that takes place in our model has no counterpart in a Cobb-Douglas economy.

\section{Stability conditions for model networks}

Here and below we will for simplicity focus on the Leontief case, commenting on the more general case $q > 0$ in the conclusion. In order to gain some intuition on the stability conditions, let us first consider a random directed network, where each supply link $J_{ij}$ is equal to $J$ with probability $r$ and $0$ with probability $1-r$, and where all $N$ firms have the same productivity $z$. The spectrum of $\mathbf{M}$ in this case is well known when $N \gg 1$ and $r \sim \mathcal{O}(1)$. It consists of an isolated eigenvalue $\lambda_{\min} = z - JrN$ and a ``sea'' of complex eigenvalues uniformly distributed in a disc of radius $J\sqrt{r(1-r)N}$ centred at $z$ (see e.g. \cite{Girko_1985,Tao_2008}). The stability condition therefore reads $z > J r N$, i.e. productivity must be large enough for the economy to function. The most unstable eigenvector, corresponding to eigenvalue $\lambda_{\min}$, is the uniform vector $(1/\sqrt{N},\ldots,1/\sqrt{N})$. As will be clear below, this corresponds to a case where crises are system wide. The same qualitative result holds when productivities are weakly heterogeneous, i.e. $z_i = z (1 + \epsilon_i)$ with $\epsilon_i \ll 1$ (albeit $\lambda_{\min}$ is slightly shifted downwards by an amount $O(z^2\epsilon^2/JN)$).

More interesting --- but more complex! --- is the case where the average number of suppliers $c = rN$ is of order unity (i.e. when $r = \mathcal{O}(N^{-1})$). In the {\it random regular network} (RRN)  where each firm has exactly $c$ suppliers (and $c$ customers) chosen randomly among the $N-1$ other firms, one knows that the spectrum of $\mathbf{M}$ again consists of an isolated eigenvalue $\lambda_{\min} = z - Jc$ and a ``sea'' of complex eigenvalues distributed in a disc of radius $J\sqrt{c}$ centred at $z$.\footnote{In this case, however, the density of complex eigenvalues is not uniform but is given by $\rho(\lambda) \propto (c^2 - |\lambda|^2)^{-2}$ for $|\lambda| < \sqrt{c}$~\cite{metz2018spectra}.} When heterogeneity is introduced, either topological (i.e. letting the number of suppliers/customers to vary) or because the couplings $J$ and the productivities $z$ fluctuate, there is no exact results available, in particular in the case where $\mathbf{M}$ is not a symmetric matrix --- see \cite{metz2018spectra} for a very recent survey. 

When $\mathbf{M}$ is symmetric, exact results are still scarce but a huge amount of work has been done in the physics, mathematics and computer science literature to characterize the eigenvalues and eigenvectors of such random matrices~\cite{wigner1967random, biroli1999single,farkas2001spectra,albert2002statistical,rogers2009cavity,fyodorov1991localization,Kuhn_2008,Benaych_Georges_2014,neri2016eigenvalue,biroli2010anderson}.
The reason is that such symmetric random matrices appear is many physical situations, such as the vibration spectrum of amorphous solids or the energy spectrum of quantum systems with impurities. Such random matrices also appear in graph theory and computer science. The problem of estimating the extremal eigenvalue is of special importance, as it appears in many different problems (such as epidemic or rumor spreading~\cite{Chakrabarti_2008} --- or crisis propagation as in the present work); the associated eigenvector is related to the concept of node centrality in network theory~\cite{Newman_2010}; see also~\cite{Castellano_2017} and refs. therein. 

\begin{figure}[tb]
    \centering
    \includegraphics[width=10cm]{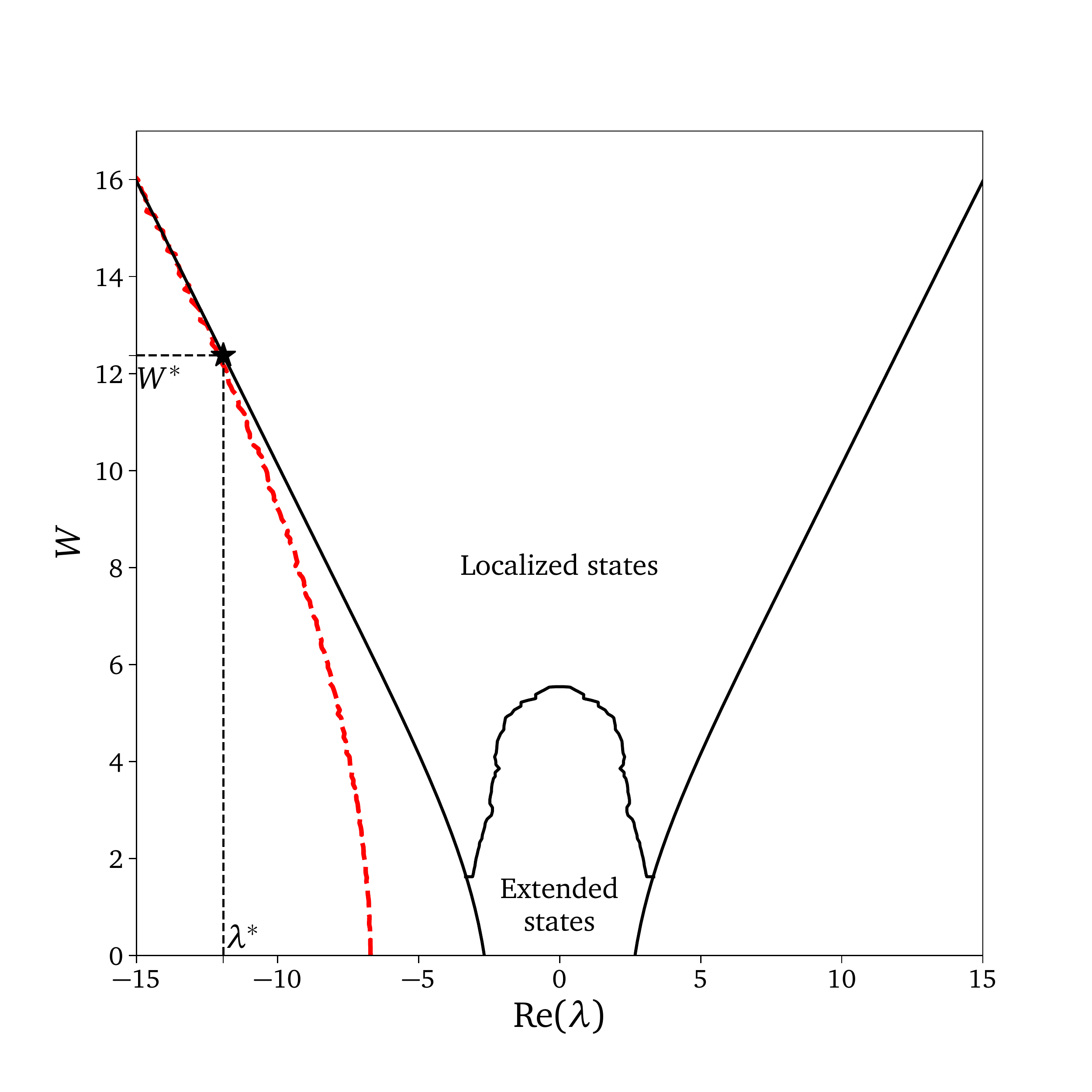}
    \caption{Numerical results for the structure of the eigenstates of a directed random regular network (RRN) with $N=2000$ firms with $c=7$ suppliers and clients each and $J = 1$. Productivities $z_i$ are uniformly distributed in an interval $[z-W/2,z+W/2]$, but the $x$-axis is centred around $z$. Notice that the eigenstates contained in the bulk get localized as $W$ increases. The left-most dashed red curve corresponds to the isolated eigenvalue that gets absorbed at the point marked by a black star in the graph, corresponding to values $W^* \approx 12.4$ and $\lambda^*\approx -12$. The boundary between extended and localized states is defined here by $H= 5/N$. Compare to Fig. 1 in~\cite{biroli2010anderson}, for the case of un-directed RRNs.}
    \label{fig:ipr_smooth}
\end{figure}

In the sequel, we will call $\Delta^2$ the variance of fluctuations (of connectivity, productivity, etc.). From the host of results accumulated in the last decades, the following general scenario is expected (see~\cite{biroli2010anderson} and Fig.~\ref{fig:ipr_smooth} for the case of random regular graphs):
\begin{itemize}
\item For $\Delta = 0$, all eigenvalues except one have their real part confined in a certain interval $\mathcal{I}$ ($=[z - \sqrt{c}, z+ \sqrt{c}]$ in the RRN example), while the isolated eigenvalue is located to the left of this interval, at a non-zero distance $g$ from its edge.    
\item As $\Delta$ increases, the interval $\mathcal{I}$ broadens and its edges become somewhat blurred, while the isolated eigenvalue gets closer and closer to the lower edge of $\mathcal{I}$ (see Fig.~\ref{fig:ipr_smooth}).
\item Beyond a certain critical value $\Delta_c$, the isolated eigenvalue is ``eaten up'' by $\mathcal{I}$ and disappears (this is called, in a different context, the Baik-Ben Arous-P\'ech\'e (BBP) transition~\cite{baik2005phase}).
\end{itemize}

Furthermore, as soon as $\Delta$ is non zero, the interval $\mathcal{I}$ is further subdivided into 3 intervals $\mathcal{I}_{-},\mathcal{I}_0,\mathcal{I}_{+}$ (with $\mathcal{I}_0$ possibly empty, see Fig.~\ref{fig:ipr_smooth}), where the structure of the corresponding eigenvectors is markedly different. In the central part $\mathcal{I}_0$, eigenvectors are {\it extended}, or {\it delocalized}, whereas in the extreme parts $\mathcal{I}_{-},\mathcal{I}_{+}$, eigenvectors are {\it localized}. In a hand-waving manner, ``localized'' means that most of the norm of the vector is concentrated on a few nodes (firms), whereas ``delocalized'' means that the norm is well spread out over all nodes. More precisely, calling $v_1,v_2,...,v_N$ the component of a normalized vector $\ket{V}$, the localized/delocalized nature of $\ket{V}$ is captured by its Herfindahl index $H$ (called Inverse Participation Ratio (IPR) in the physics literature):
\begin{equation}\label{eq:IPR_def}
H(\ket{V}) = \sum_{i}^{N} |v_i|^4.
\end{equation}
A localized eigenvector is such that $H(\ket{V}) = \mathcal{O}(1)$ in the limit $N \to \infty$ whereas a delocalized eigenvector has $H(\ket{V}) = \mathcal{O}(N^{-1})$.
The importance of this distinction for crisis propagation in the context of our model will become clear below. 

Owing to the structure of $\mathbf{M}$, the Perron-Frobenius theorem ensures that its leftmost eigenvalue $\lambda_{\min}$ is real with a real positive eigenvector $u_i > 0$. As stated above, $\lambda_{\min}$ must be positive for $\mathbf{M}$ to be an M-matrix, i.e. for all prices and all quantities to be positive. As $\lambda_{\min} \to 0$, the economy becomes more and more fragile to external shocks. Let us for example consider the case where the productivity $z_i$ of some firms decrease by $- \varepsilon \Delta_i < 0$, and/or that some of the stoichiometric weights $J_{ij}$ increase by some amount $\varepsilon \Delta_{ij} > 0$. Using standard perturbation theory to first order in  $\varepsilon$\footnote{Indeed, given an eigenvector $\vec{u}$ corresponding to an eigenvalue $\lambda$ of a matrix $\mathbf{B}$ subject to a perturbation $\mathbf{B}\rightarrow \mathbf{B}+\varepsilon\mathbf{P}$, the first order correction to $\lambda$ in epsilon reads $\lambda \rightarrow \lambda + \varepsilon ^t\vec{u}\mathbf{P}\vec{u}$.}, one finds that the leftmost eigenvalue is shifted as:
\begin{equation}\label{eq:shift}
\lambda_{\min} \longrightarrow \lambda_{\min} - \varepsilon \left[ \sum_i \Delta_i u_i^2 + \sum_{i \neq j} \Delta_{ij} u_i u_j \right]. 
\end{equation}
Since both correction terms are negative, this formula shows that as the system becomes marginally stable, any local decrease of productivity/increase of required inputs tips the system towards the unstable region. A certain number of prices/quantities then become negative. Intuitively, the physiognomy of these ``crises'' will depend on the localized/delocalized nature of the eigenvector $\ket{U}$ corresponding to $\lambda_{\min}$, as we now discuss.

The next order correction to Eq. (\ref{eq:shift}) is of order $\varepsilon^2/g$, where $g$ is the gap between $\lambda_{\min}$ and the next eigenvalue of $\mathbf{M}$; therefore first order perturbation theory is only valid provided $\varepsilon \ll g$. Now, two cases must be distinguished, depending on the strength $\Delta$ of the heterogeneities:
\begin{itemize}
\item When $\Delta < \Delta_c$, the leftmost eigenvalue is isolated, in which case $g > 0$ even when $N \to \infty$. The first order result Eq. (\ref{eq:shift}) is then valid when $\varepsilon$ is small enough. Furthermore, the associated eigenvector $\ket{U}$ is delocalized. From Eq. (\ref{eq:shift}), one deduces that a localized shock --- say on firm $\ell$ alone --- decreases $\lambda_{\min}$ by $\sim -\delta z_\ell/N$. The system is unstable when $\delta z_\ell > N \lambda_{\min}$, but for this condition to be compatible with $\delta z_\ell \ll g$, one must also require $N \lambda_{\min} \ll g$. When destabilized, the shock propagates over the whole system, because of the delocalized nature of $\ket{U}$. In the case of a small {\it global} productivity shock $\delta z_i = \delta z, \forall i$, the destabilisation occurs as soon as $\delta z > \lambda_{\min}$. 
\item When $\Delta > \Delta_c$, the leftmost eigenvalue is at the edge of the interval $\mathcal{I}_-$, such that the gap $g(N)$ generically goes to zero as $N \to \infty$.  Furthermore, the associated eigenvector $\ket{U}$ is now localized, usually centred around particularly low productivity/high connectivity firms (called the Lifschitz regions in the physics literature~\cite{Lifshitz_1964,Thouless_1974,biroli2010anderson}). In this case, however, first order perturbation theory breaks down as soon as $\varepsilon \sim g(N)$, so one must have recourse to numerical simulations to characterize the associated crisis patterns --- see next section and Appendix~\ref{sec:real_networks} for a comparison with empirical data.  
\end{itemize} 

Finally, the following remarks should be useful to get an intuition about crisis propagation in our model. Note that one can write prices and outputs using the inverse matrix $\mathbf{M}^{-1}$ and its transpose. Hence, the price response to some generic perturbations $\delta y$ (for example to productivity, household consumption, etc.) can be expressed using the eigenvalues and eigenvectors as:
\begin{equation}\label{eq:m_inv_eigen1}
\delta p_i = \sum_{\alpha} \ell_i^\alpha \frac{1}{\lambda_\alpha} \braket{r^\alpha \vert \delta y},
\end{equation}
where $\ell^\alpha, r^\alpha$ are, respectively, the left and right eigenvectors of $\mathbf{M}$ associated to eigenvalue $\lambda_\alpha$. Similarly, for production
\begin{equation}\label{eq:m_inv_eigen2}
\delta \gamma_i = \sum_{\alpha} r_i^\alpha \frac{1}{\lambda_\alpha} \braket{\ell^\alpha \vert \delta y}.
\end{equation}
In the limit where $\lambda_{\min}$ touches zero with a finite gap $g$, one can approximate these responses as
\begin{equation}\label{eq:m_inv_eigen3}
\delta p_i \approx \frac{\ell_i^{\min} \braket{r^{\min} \vert \delta y}}{\lambda_{\min}}; \qquad \delta \gamma_i \approx \frac{r_i^{\min} \braket{\ell^{\min} \vert \delta y}}{\lambda_{\min}}.
\end{equation}
Hence, the amplitude of the response of prices depends on the overlap $\braket{r^{\min} \vert \delta y}$ and is localized on the left eigenvector $\ell^{\min}$, and vice-versa for production. This will be illustrated using real data in Appendix \ref{sec:real_networks}. 

In order to understand intuitively the divergence of the response to perturbations, consider the simple case where $\forall i,\, z_i=z$. One can expand $\mathbf{M}^{-1}$ in the stable region as 
\begin{equation}\label{eq:m_inv}
\mathbf{M}^{-1}=\frac{1}{z}\sum_{k=0}^{\infty}\left(\frac{\mathbf{J}}{z}\right)^k
\end{equation}
with $(\mathbf{J})_{ij}=J_{ij}$, since the stability condition implies that the spectral radius of $\mathbf{J}$ is smaller than $z$. Now, the term $(\mathbf{J}^k)_{ij}$ consists of the sum of all paths of length $k$ linking firm $j$ to firm $i$. Marginal stability corresponds to this sum becoming divergent, with paths of all lengths contributing to $(\mathbf{M}^{-1})_{ij}$. This interpretation also holds in case of heterogeneous $z_i$s.\footnote{Indeed one can always write an M-matrix as $\mathbf{M}=z_{\max}\mathbf{1}-\mathbf{B}$ where $\mathbf{B}$ is non negative, and expand the series as $\mathbf{M}^{-1}=\frac{1}{z_{\max}}\sum_{k=0}^{\infty}\left(\frac{\mathbf{B}}{z_{\max}}\right)^k$.} The instability is therefore related to a situation where shocks can propagate over paths of arbitrary length in the input-output network. This is closely related to second-order phase transitions in physics, where correlations extend over macroscopic distances and the response to small perturbations diverges, see e.g.~\cite{sethna2006statistical}. 

\section{Numerical results: broad distribution of firm sizes and crises}

We will consider the simplest model of a random regular network, where each firm has exactly $c$ suppliers and $c$ customers, each chosen randomly among the $N-1$ other firms (other types of networks will be discussed in~\cite{ustocome}). Each firm has a random productivity uniformly chosen in the interval $[z - W/2, z + W/2]$, such that $z$ is the average productivity and $\Delta = W/2\sqrt{3}$. The stoichiometric coefficients $J_{ij}$ are taken to be all equal to $J$. Without loss of generality, $J$ can be set to unity. In addition, we set $V_i=1$ for simplicity and take the households' consumption $C_i=1/p_i$ as obtained from a logarithmic utility function with identical preference for all products.   

\begin{figure}[tb]
    \centering
    \includegraphics[width=10cm]{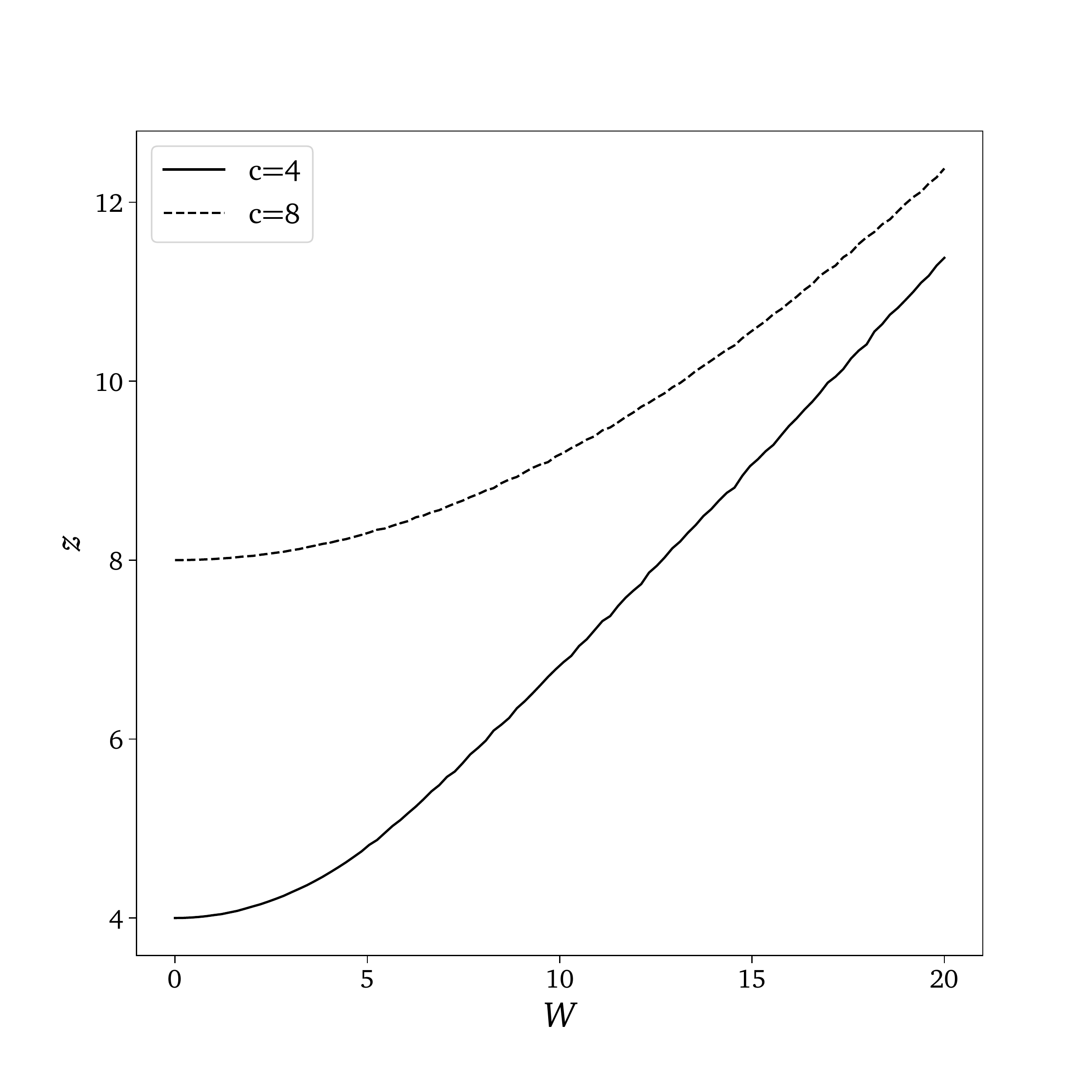}
    \caption{Plot of the average productivity $z=f_c(W)$ needed to stabilize the economy in the case of a directed RRN with productivities uniformly distributed in $[z-W/2; z+W/2]$, for connectivities $c=4$ and $c=8$. Notice that we find $f_c(0)=c$ and a linear behaviour of $f_c(W)$ as $W\to\infty$, as expected. Economies with $N=2048$ firms were simulated. Error bars are too small to be visible.}
    \label{fig:fofw}
\end{figure}

When $W=0$, the spectrum of $\mathbf{M}$ can be computed, as discussed in the previous section, with an isolated leftmost eigenvalue given by $\lambda_{\min} = z - Jc$. As $W$ increases, the spectrum evolves as shown in Fig.~\ref{fig:ipr_smooth}. In the case of $c=7$,  the isolated eigenvalue disappears when $W=W^* \approx 15.5$, and the edge of the spectrum corresponds to a localized eigenvector. In the following, we fix the average productivity to $z=f_c(W)$ such that $\lambda_{\min} = \epsilon = 10^{-8}$ (the function $f_c(W)$ is shown in Fig.~\ref{fig:fofw}). The model then only depends on two parameters: the connectivity $c$ and the productivity heterogeneity $W$. We will study along this critical line different characteristics of the corresponding economy. 

\begin{figure}[tb]
    \centering
    \includegraphics[width=10cm]{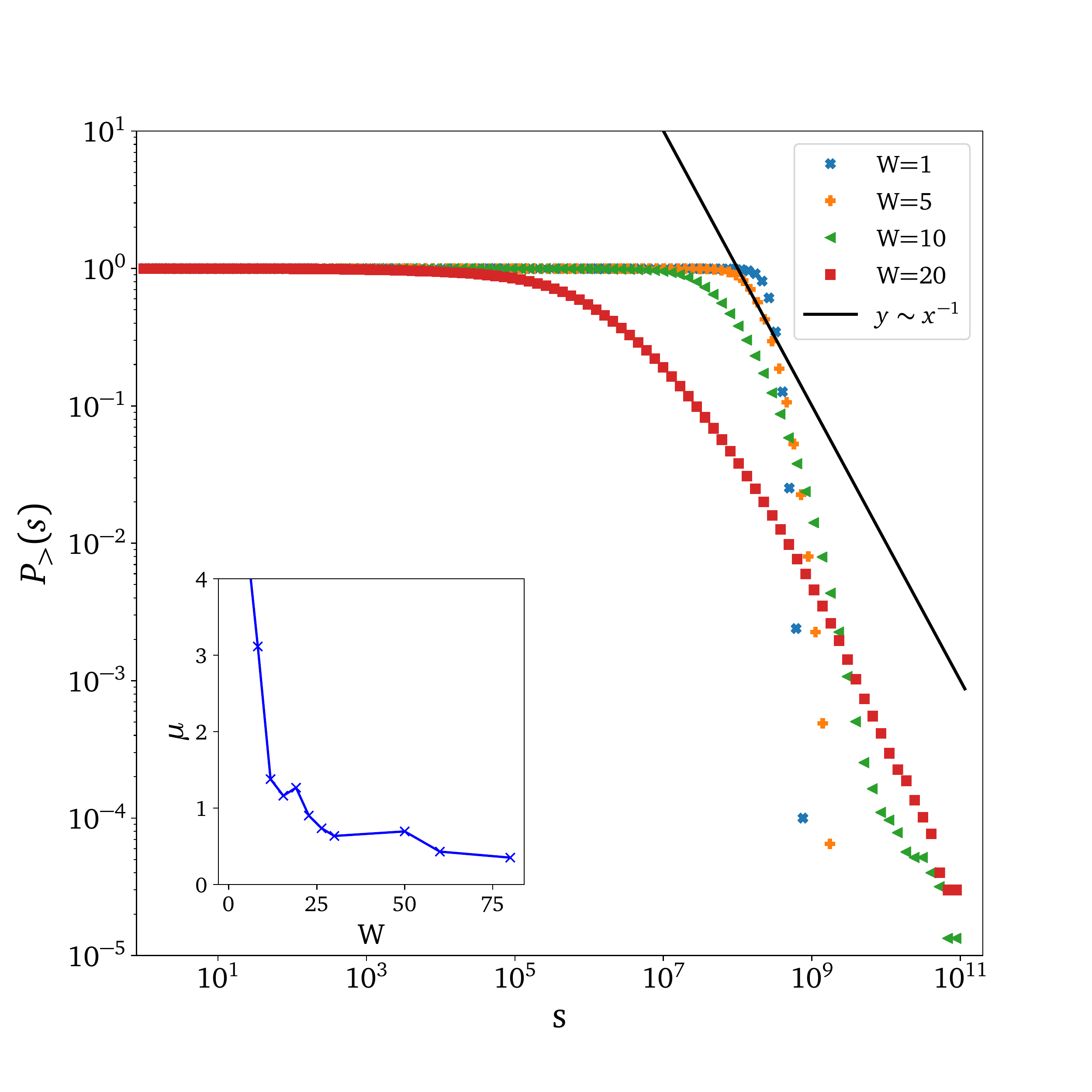}
    \caption{Log-log cumulative distribution of firm sizes $P_>(S)$ defined by sales $S_i=z_i\gamma_i p_i$, along with a curve corresponding to power-law $S^{-\mu}$ with exponent $\mu=1$ (Zipf) for comparison. For small values of $W$ the distribution of firm sizes is sharply peaked at a value of order $1/\varepsilon = 10^8$. Increasing $W$ causes the distribution to get fatter tails with an apparent power-law exponent $\mu$ that decreases with $W$. Here, $c=4$, $z=f_c(W)$, $N=1500$. Inset: power-law exponent $\mu$ as a function of $W$, along the critical line of the model.}
    \label{fig:firm_size_cum}
\end{figure}

Two quantities are of particular interest for this paper (a more throughout account of the results will be reported in~\cite{ustocome}). One is the distribution of firm size, defined as the total sales $\mathcal{S}_i=z_i \gamma_i p_i$. Quite interestingly, while this distribution has thin tails when $W$ is small, it becomes fat-tailed (Zipf-like) as $W$ increases, as observed empirically~\cite{Axtell_2001}, but with an exponent that appears to vary with $W$ and $c$ (see Fig.~\ref{fig:firm_size_cum}, inset).
The emergence of a power-tailed firm size distribution is a consequence of the criticality of the model, but requires the extreme eigenvectors to be localized and heterogeneous, as it is the case for $W$ sufficiently large.

\begin{figure}[tb]
    \centering
    \includegraphics[width=10cm]{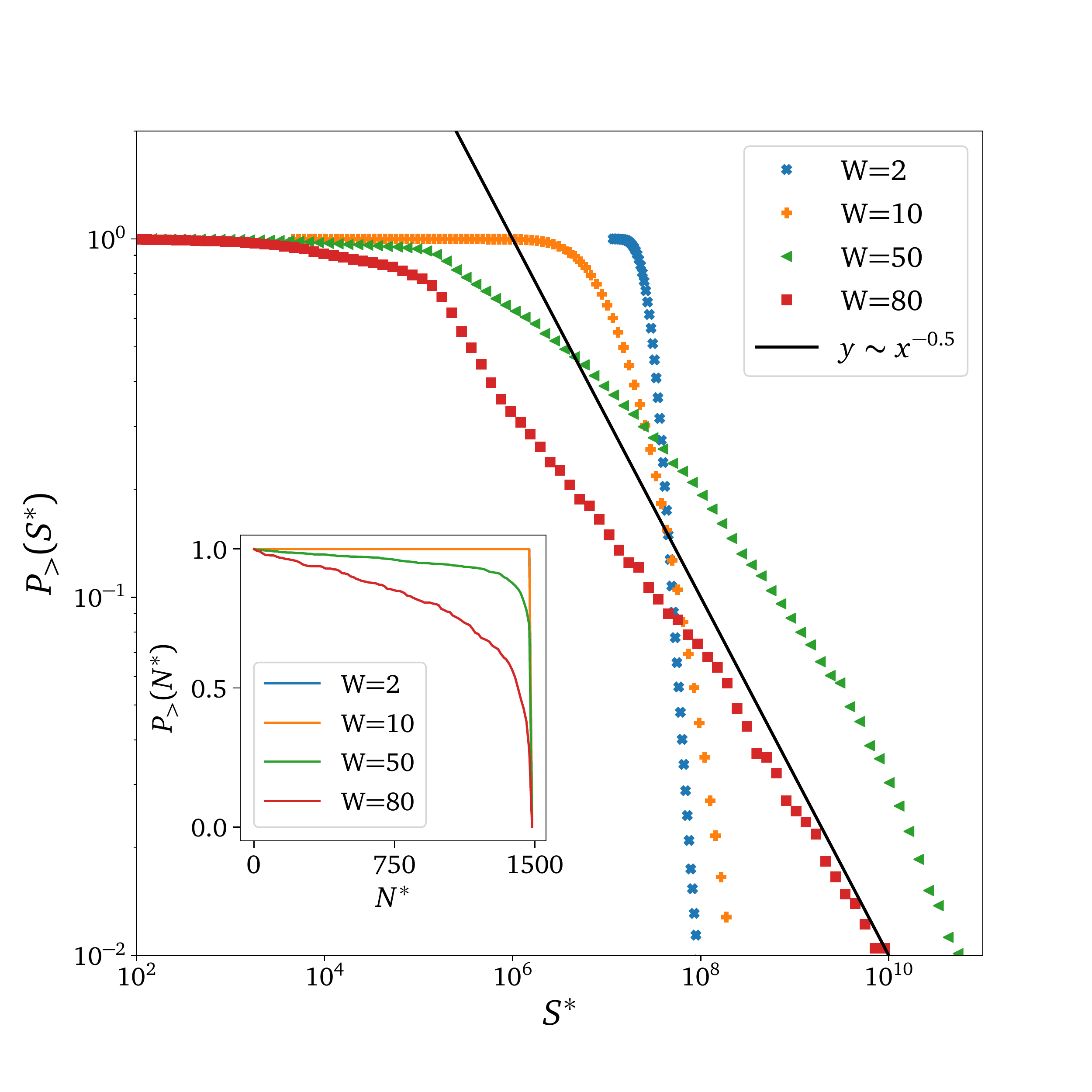}
    \caption{Log-log cumulative distribution of avalanche sizes as defined by the total sales of firms gone under after a shock, for $c=4$, $z=f_c(W)$, $N=1500$ and $\delta z_\ell = 0.05$. One again observes a broad, power-law tailed distribution of casualties for large enough $W$'s. Inset: cumulative distribution of the number of firms $N^*$ that have gone under after a shock. Notice that $W=2$ and $W=10$ correspond to mostly system-wide avalanches (i.e. $N^* \approx 1500$), while larger values of $W$ correspond to avalanches of all sizes. }
    \label{fig:avalanches}
\end{figure}

The second quantity is the distribution of crises amplitudes $\mathcal{A}$, defined as the total size of the firms that are such that their equilibrium price becomes negative after an idiosyncratic shock of amplitude $-\delta z_\ell$ hitting a certain firm $\ell$. (Shocks on the coefficients $J_{ij}$ lead to qualitatively similar results). While in a fraction of cases nothing much happens, an avalanche can develop where a number of firms ``go under'', in the sense that their equilibrium price becomes negative. Conditioned to such events, the distribution of $\mathcal{A}$ is found to be of three types (see Fig~\ref{fig:avalanches}):
\begin{enumerate}
\item  mostly ``system wide'', where a substantial fraction of the output is wiped out. This occurs when $W < W^*$ and  $\delta z_\ell > N \lambda_{\min}$, as expected from our general discussion;
\item  thin-tailed, where avalanches are restricted to particularly fragile firms connected to $\ell$. This corresponds to $W > W^*$, and weak perturbations $\delta z_\ell < g$, in which case only one or a few localized 
eigenvectors close to the edge propagate the crisis.
\item fat-tailed, where small crises coexist with large crises (a feature of Self-Organized Criticality, as recalled in the introduction). This happens when $\delta z_\ell \gg g$, i.e. when a large collection of eigenstates are mobilized in the crisis propagation.
\end{enumerate}
The generic existence of three crisis scenarii is, we believe, quite interesting. In particular, the possibility that a small, idiosyncratic shock can lead to system-wide trouble, or else to avalanches of all sizes, has potentially deep consequences on our understanding of the business cycle and on crisis prevention policies. Of course, the above analysis postulates that the economy is close to criticality, i.e. that $\lambda_{\min} \to 0$. Why this should be the case is obviously the crux of the matter, and will be discussed in the next section. 

\begin{figure}[tb]
    \centering
    \includegraphics[width=10cm]{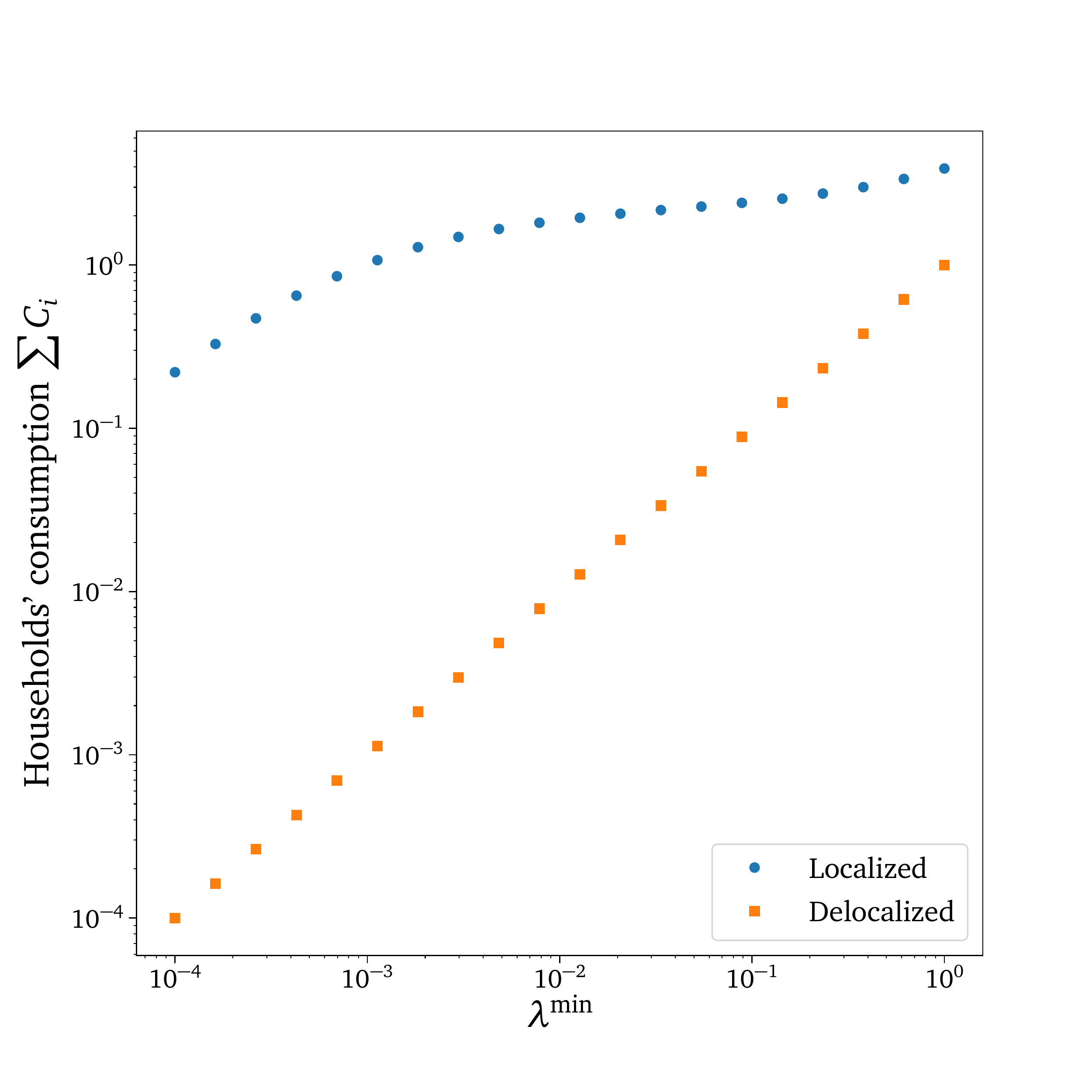}
    \caption{Total consumption of households vs. $\lambda_{\min}$ in the localized and delocalized cases. Intuitively, one expects the prices of goods represented in the eigenvector corresponding to $\lambda_{\min}$ to behave as $\lambda_{\min}^{-1}$, while a logarithmic utility function implies that consumption scales as the inverse of the price, leading naturally to the consumption of those goods to be proportional to $\lambda_{\min}$. In the delocalized case, all goods are concerned and thus global consumption plummets, as seen in the graph (orange squares). On the other hand, only the handful of goods associated to $\lambda_{\min}$ see their consumption decline in the delocalized case, corresponding to the blue dots. An economy with $1000$ firms and a connectivity $c=4$ was used for this plot, with $W=0$ for the localized case and $W=12$ for the localized case.}
    \label{fig:consumption}
\end{figure}

It is also interesting to plot the total consumption of households as a function of $\lambda_{\min}$ in the two cases above: delocalized ($W < W^*$) vs. localized ($W < W^*$) crises. In the first case, the whole economy grinds to a halt as $\lambda_{\min} \to 0$, as expected. In the second case, only a fraction of the total consumption (mostly coming from firms represented in the corresponding localized eigenvector) is affected. See Figure~\ref{fig:consumption} for an illustration of this point.

Our general scenario is strongly reminiscent of similar ideas in an ecological context, where the disappearance of a single species can lead to mass extinctions mediated by network effects~\cite{may1972will,biroli2018marginally}. A major difference, however, is that the economic network is not static and can in principle adapt to new conditions on relatively short time scales (at least compared to evolutionary timescales). With our framework, we expect that if a supplier undergoes some difficulty (i.e. its equilibrium production is found to be negative), its customers may choose to rewire and look for alternatives. Furthermore, one expects that the market clearing and zero profit conditions will be temporarily violated. An extension of the present model that takes such dynamical effects into account would certainly be extremely interesting (see~\cite{bonart2014instabilities, colon2019} for preliminary work in that direction). But what is clear is that if rewiring takes time and/or is costly, the ``paper crises'' found above could indeed materialize as actual defaults, or at least acute difficulties. Since economic frictions are substantial and rewiring cannot be instantaneous, we expect that the present scenario could be relevant to understand real world crises~\cite{reed_simon_2011,reuters_japan_2011,fisher_hbs_2011}. 

Let us finally come back to the case of partial substitutability, i.e. when the parameter $q$ appearing in Eq. (\ref{eq:def_ces}) is strictly larger than zero. Since the matrix  
$(\widehat{\mathbf{M}})_{ij}=z_i^\zeta \delta_{ij}- a_{ij}^{q \zeta} J_{ij}^\zeta$ is a continuous function of $q$, it is clear that if the smallest eigenvalue $\lambda_{\min}$ is negative for $q=0$, it will remain so for a certain range of $q$. We have checked numerically on some examples that this is indeed the case; the economy is only stabilized when $q$ exceeds a (problem dependent) value $q_c > 0$. Not surprisingly, allowing for more substitutability can stabilize an otherwise unfeasible economy. An in-depth study of this new threshold will be presented in our forthcoming work~\cite{ustocome}, but we expect all the properties reported in this section to hold true for all $q > 0$ when the system is close to criticality.

\section{Marginal stability: Discussion \& Conclusion} 
\label{sub:marginal_stability}

In this section, we will motivate our claim that generic economies --- like many other complex systems, see e.g.~\cite{bak2013nature,le2010avalanches,charbonneau2014fractal,muller2015marginal,biroli2018marginally} --- might ``self-organize'' to sit, at least temporarily, close the boundary of the stable region, i.e. satisfy the marginal stability criterion $\lambda_{\min} \to 0$. Several types of evolutionary forces act to that effect. One is simply the creation of new firms, that lead to an {\it effective} reduction of productivity and increase of connectivity. To see this, consider that the economy  consists of $N$ firms in equilibrium and add an additional firm indexed by $\star$, with productivity $z_\star$, labour requirements $J_{\star0}=V_\star/p_0$ and links $J_{\star i},J_{j\star}$ to the $N$ pre-existing firms. The equilibrium condition for price $p_\star$ is:
\begin{equation}\label{eq:price_addition}
p_\star =  \frac{V_\star}{z_\star} + \sum_{j=1}^{N}\frac{J_{\star j}}{z_\star}p_j. 
\end{equation} 
Plugging this result in the new equilibrium conditions for the $N$ original firms yields:
\begin{equation}\label{eq:renormalized}
(z_i-\frac{J_{i\star}J_{\star i}}{z_\star}) p_i - \sum_{j=1}^{N} \left(J_{ij}+\frac{J_{i\star}J_{\star j}}{z_\star}\right)p_j = V_i+\frac{J_{i\star}V_\star}{z_\star}
\end{equation}
which means that the addition of a firm amounts in effect to {\it decreasing} all original productivities: $z_i \to z_i-\frac{J_{i\star}J_{\star i}}{z_\star}$ and {\it increasing} all stoichiometric coefficients: $J_{ij}\to J_{ij}+{J_{i\star}J_{\star j}}/{z_\star}$.  
As clear from Eq. (\ref{eq:shift}), this can only decrease the smallest eigenvalue of the matrix $\mathbf{M}_N^\star$ that describes the pre-existing firms with the new firm added. One concludes that a growing economy can only become more unstable with time. This argument is actually closely related to May's original argument about the stability of large ecologies~\cite{may1972will}. 

In fact, one can show that as the number of links to the most connected node of the network increases, the smallest eigenvalue of $\mathbf{M}$ decreases \cite{Castellano_2017}, until the instability threshold is reached. In this case, the fragility of the network comes from the most central hubs, a scenario akin to, but different from, the one of Acemoglu, Carvalho et al.~\cite{acemoglu2012network}. This effect might be amplified if firms systematically favour links toward hubs (as suggested in~\cite{fisher_hbs_2011}), leading to a ``scale-free'' input-output network~\cite{atalay2011}. Interestingly, a stability-constrained growth mechanism for networks, whereby a node is freely added to the network if it does not destabilize the system but induces rewirings in the network until stability is found again if it does, has been found to generate such scale-free networks~\cite{Perotti2009}.

The second evolutionary effect is, even for a fixed size $N$, the complexification of the production process, i.e., technology progress means that a wider array of products are needed as inputs. If the average productivity $z$ remains the same while the average connectivity $c$ increases, the system eventually reaches the instability point (which in the simplest case reads $z = Jc$). Hence productivity must increase at some minimum rate for the economy to remain stable. But since increasing productivity is costly, one can postulate that the average productivity $z$ will tend to hover around the minimal viable threshold, and sometimes lagging behind, leading to occasional endogenous crises. Similarly, as mentioned after Eq. (\ref{eq:def_leontieff}), firms tend to optimize their portfolios of suppliers, thereby reducing their redundancy but, by the same token, reducing the effective substitution effects captured by the CES parameter $q$. As $q \to q_c^+$, the economy will again become unstable. 

Finally, we have assumed that firms are perfectly competitive and that equilibrium corresponds to zero profit. Now, in more realistic situations, firms attempt to realize positive profits and distribute dividends. As already noted, if the profit target of firm $i$ is a certain fraction $\varphi_i$ of its total sales $S_i=z_i \gamma_i p_i$, Eq. (\ref{eq:excess_prof}) remains identical but with a decreased effective productivity $z_i \to z_i(1-\varphi_i)$. As firms attempt to maximize their profits, the average effective productivity goes down, until the marginal stability point is reached and a crisis ensues. After the crisis, economic actors revert to more reasonable levels of markups (i.e. reduce $\varphi_i$), which makes the economy viable again --- until the next crisis.  

One could probably come up with other mechanisms that push the economy towards instability, see for example \cite{demartino_2007,Bardoscia_2017}. Our conjecture is that evolutionary and behavioural forces repeatedly drive the economy close to marginal stability. As anticipated by Bak, Chen, Scheinkman, and Woodford~\cite{bak2013nature, bak1993aggregate,scheinkman1994self} and confirmed in this paper, this scenario would be a natural explanation of the broad (Zipf-like) distribution of firm sizes, and of the ``small shocks large business cycle'' puzzle, that both suggest some kind of criticality. Crises should then be understood as intrinsically non-linear events, where feedback loops of arbitrary size contribute to propagating and amplifying idiosyncratic shocks. 

There are many directions to explore further. The most important one is, in our opinion, to endow the model with some realistic dynamics, partly along the lines of \cite{bonart2014instabilities}, that would include frictions, myopia, imperfect market clearing, rewiring, etc. This would make the model more realistic, and is a prerequisite to calibration it on empirical data, since within the present static setting crises are signaled by the appearance of negative prices, beyond which the model ceases to make sense.

\section*{Acknowledgements} 

We are indebted to F. Benaych-Georges, G. Biroli, J. Bonart, G. Bunin, C. Colon, R. Farmer, A. Kirman, A. Landier, A. Mandel, M. Marsili, J.P. Nadal, F. Roy, A. de Sanctis, A. Secchi, D. Sharma, M. Tarzia, D. Thesmar and F. Zamponi for countless illuminating discussions on the topics of this paper. We thank in particular X. Gabaix for many detailed comments, and for insisting that we should investigate the model beyond its Leontief limit. This paper is dedicated to the memory of Per Bak, who disappeared much too early but whose seminal ideas are still extremely vivid and influential.

\appendix
\section{CES Production Functions}
\label{sub:appendix}
Of standard usage in economics, the constant elasticity of substitution (CES) functions are a family of production functions giving the total production of a firm $i$ given inputs $Q_{ij}$ from its suppliers \footnote{This may also include labour inputs, which we conventionally choose to correspond to the index $j=0$.}. In the most general setting, the CES production function is defined as
\begin{equation}\label{eq:ces_prod}
 \pi_i = z_i \left(\sum_{j}a_{ij}\left(\frac{Q_{ij}}{J_{ij}}\right)^{-1/q}\right)^{bq}
 \end{equation} 
where $z_i$ is the productivity level of firm $i$, the $a_{ij}$s are weight coefficients satisfying $\sum_ja_{ij}=1$ and the $J_{ij}$ terms are stoichiometric coefficients defining the number of inputs from $j$ required to make an unit of $i$'s output. In the remaining terms, $b$ is called the returns to scale: multiplying all inputs $Q_{ij}$ by some coefficient $K$ will make the whole output level to be multiplied by $K^b$. In the main body and all that follows we have chosen $b=1$, corresponding to the so-called constant returns to scale case, but our analysis can be extended to other values of $b$. The effect of $q$, the degree of substitutability, deserves a more in-depth discussion through the study of the limits $q\to \infty$ and $q\to0$, corresponding to the so-called Cobb-Douglas and Leontief production functions.    

\paragraph{Perfect substitutability: Cobb-Douglas case} take indeed the limit $q\to\infty$ as

\begin{equation}\label{eq:cobb-doug-case}
\begin{split}
\pi_i /z_i&= \exp\left(-q\log\left(\sum_j a_{ij}\exp\left(-\frac{1}{q}\log\left(\frac{Q_{ij}}{J_{ij}}\right)\right)\right)\right)\\
&\simeq \exp\left(-q\log\left(1-\frac{1}{q}\sum_j a_{ij}\log\left(\frac{Q_{ij}}{J_{ij}}\right)\right)\right)\\
&\simeq \exp\left(\sum_j a_{ij}\log\left(\frac{Q_{ij}}{J_{ij}}\right)\right)=\prod_j \left(\frac{Q_{ij}}{J_{ij}}\right)^{a_{ij}}
\end{split}
\end{equation}
and one retrieves the ubiquitous Cobb-Douglas production function. In this setting, one can easily check that if a given input $Q_{ij}$ from a firm $j$ is multiplied by an amount $f<1$ then the output need not drop if any other input any other input $Q_{ik}$ is multiplied by by $f^{-\frac{a_{ik}}{a_{ij}}}$.

\paragraph{Unsubstitutable inputs: Leontief case} take instead the limit $q\to 0$, and consider $j^*=\argmin_j \left(\frac{Q_{ij}}{J_{ij}}\right)$ to get

\begin{equation}\label{eq:leontief-case}
 \begin{split}
  \pi_i/z_i =& \left(a_{ij^*}\left(\frac{Q_{ij^*}}{J_{ij^*}}\right)^{-\frac{1}{q}}+\sum_{j\neq j^*}a_{ij^*}\left(\frac{Q_{ij}}{J_{ij}}\right)^{- \frac{1}{q}}\right) ^{-q}\\
  =& \frac{Q_{ij^{*b}}}{J_{ij^{*b}}}\left(a_{ij^*}+\sum_{j\neq j^*}a_{ij}\left(\frac{J_{ij^*}Q_{ij}}{Q_{ij^*}J_{ij}}\right)^{-\frac{1}{q}}\right)^{-q}\\
  \underset{q\to0}{\longrightarrow}&\frac{Q_{ij^{*}}}{J_{ij^{*}}}=\min_j \left(\frac{Q_{ij}}{J_{ij}}\right)
 \end{split}
\end{equation} 
where the total output of firm $i$ is determined by its scarcest input.

The CES production function therefore bridges these two limiting cases of which only the Cobb-Douglas case has been studied in the network literature. We will now study the competitive equilibrium equations for values of $q\in[0;\infty)$.

\section{Competitive Equilibrium Equations and Hawkins-Simon Condition} 
\label{sub:competitive_equilibrium_equations_and_hawkins_simon_condition}
Our problem is first to maximize the profits
\begin{equation}\label{eq:profits}
    \mathcal{P}_i = \pi_i p_i - \sum_j Q_{ij}p_j
\end{equation}
for each firm subject to the constraint given by eq.\eqref{eq:ces_prod}. Computing first the derivative of $\pi_i$ w.r.t. $Q_{il=j}$ and substituting using eq.\eqref{eq:ces_prod} yields
\begin{equation}
    \frac{\partial \pi_i}{\partial Q_{ij}} = z_i^{-\frac{1}{q}}a_{ij}J_{ij}^{\frac{1}{q}}Q_{ij}^{-\frac{1+q}{q}}\pi_i^{\frac{1+q}{q}}
\end{equation}
which can now be used to set $\frac{\partial \mathcal{P}_i}{\partial Q_{ij}}=0$, i.e.
\begin{equation}
\begin{split}
p_i z_i^{-\frac{1}{q}}a_{ij}J_{ij}^{\frac{1}{q}}Q_{ij}^{-\frac{1+q}{q}}\pi_i^{\frac{1+q}{q}} &= p_j\\
Q_{ij} &= \left(\frac{p_i}{p_j}a_{ij}\right)^{\frac{q}{1+q}}J_{ij}^{\frac{1}{q+1}}z_i^{-\frac{1}{1+q}}\pi_i.
\end{split}
\end{equation}
One can now check this solution in the Leontief and Cobb-Douglas limiting cases
\begin{equation}\label{eq:leontief-cd-limit}
Q_{ij} \left\{ \begin{matrix}
&\underset{q\to\infty}{=}&\frac{p_i}{p_j}a_{ij} \pi_i\\
&\underset{q\to0}{=}& J_{ij} \frac{\pi_i}{z_i}
\end{matrix}\right.
\end{equation}
retrieving the condition $Q_{ij}=J_{ij}\gamma_i$ defined in eq.\eqref{eq:output_def} of the main body, where $\gamma_i:=\pi_i/z_i$ is the firm's output level. 

In the Leontief case the optimal input is necessarily determined by the firm's desired output level, while it is determined by the input's price in the Cobb-Douglas case.

One needs now to impose a competitive equilibrium by setting all optimized profits to $0$ as
\begin{equation}\label{eq:competitive_eq_q}
\begin{split}
\pi_i p_i &= \sum_j Q_{ij}p_j\\
\pi_i \left(z_i^{\frac{1}{1+q}}p_i^{\frac{1}{1+q}}\right) &= \pi_i\left( \sum_j a_{ij}^{\frac{q}{1+q}}J_{ij}^{\frac{1}{1+q}}p_j^{\frac{1}{1+q}}\right)
\end{split}
\end{equation}
corresponding to eq.\eqref{eq:excess_prof2}. The existence of a solution $p_i>0$ for this equation is equivalent to saying that the matrix $(\widehat{\mathbf{M}})_{ij}=z_i^\zeta \delta_{ij}- a_{ij}^{q \zeta} J_{ij}^\zeta$, with $\zeta = 1/(1+q)$, is a so-called M-matrix.
Once prices are determined, imposing market clearing $\pi_i = \sum_j Q_{ji}$ also leads to a similar equation
\begin{equation}\label{eq:q_mclearing}
\begin{split}
z_i\gamma_i - \sum_j \left(\frac{p_j}{p_i}a_{ji}\right)^{q\zeta}J_{ji}^\zeta z_j^{q\zeta}\gamma_j&= Q_{j0}
\end{split}
\end{equation}
which would require the matrix $\left(\widetilde{\mathbf{M}}\right)_{ij}=z_i\delta_{ij}-\left(\frac{p_j}{p_i}a_{ji}\right)^{q\zeta}J_{ji}^\zeta z_j^{q\zeta}$ to be an M-matrix. In the Leontief $q\to 0$ case we have the relation $\widetilde{\mathbf{M}}= \widehat{\mathbf{M}}^\intercal$ and the sufficient condition to have both a competitive zero profit equilibrium and market clearing is that $\widehat{\mathbf{M}}$ be an M-matrix. 

On the other hand, in the Cobb-Douglas case, it is easy to see that $(\widehat{\mathbf{M}})_{ij}=\delta_{ij}- a_{ij} $ is always an M-matrix, and thus the prices are chosen so that  $\left(\widetilde{\mathbf{M}}\right)_{ij}=z_i\delta_{ij}-\frac{p_j}{p_i}a_{ji}z_j$ is an M-matrix to have a market-clearing equilibrium.

\section{Functional Economies}
\label{sub:appendixB}
In \cite{hawkins1948some,hawkins1949note} the authors proved that a necessary and sufficient condition for an equation such as eq.\eqref{eq:excess_prof} to have solutions is that all of the principal minors of the matrix $\mathbf{M}$ be positive, while the authors in \cite{fiedler1962matrices} proved that this condition is equivalent to all of the eigenvalues of $\mathbf{M}$ have a positive real part. 

Regarding the positivity of all principal minors of $\mathbf{M}$, Hawkins and Simon gave an economic interpretation of this condition by claiming that it is equivalent to saying that ``the group of industries corresponding to each minor must be capable of supplying more than its own needs for the group of products produced by this group of industries''\cite{hawkins1949note}. In this section we will provide a rewording of this in terms of an effective medium equation/Schur complements.

\subsection{Effective medium for one firm} 
\label{sub:effective_medium_for_one_firm}

Consider first a firm $i$ satisfying eq.\eqref{eq:excess_prod} written as
\begin{equation}
    \mathbf{M}\ket{P} = \ket{V}
\end{equation}
and consider the matrix $\mathbf{M}^{(i)}$ with row and column $i$ removed, as well as $\ket{P^{(i)}}$ and $\ket{V^{(i)}}$ the vectors with $i$ removed, $\ket{J_{i\leftarrow}}=\left(J_{i0},\ldots,J_{iN}\right)$ and $\ket{J_{i\rightarrow}}= \left(J_{0i}\ldots,J_{Ni}\right)$. The previous equation can now be written as
\begin{equation}\label{eq:schur_1firm}
\begin{split}
z_i p_i - \braket{J_{i\leftarrow}\vert P^{(i)}} &= V_i \\
\mathbf{M}^{(i)}\ket{P^{(i)}} - p_i \ket{J_{i\rightarrow}} &= \ket{V^{(i)}} 
\end{split}
\end{equation}
multiplying now the second line by $\left(\mathbf{M}^{(i)}\right)^{-1}$, taking the product with $\bra{J_{i\leftarrow}}$ and subtracting from the first line leads to
\begin{equation}
    \left(z_i - \braket{J_{i\leftarrow}\vert(\mathbf{M}^{(i)})^{-1}\vert J_{i\rightarrow}}\right)p_i = V_i + \braket{J_{i\leftarrow}\vert(\mathbf{M}^{(i)})^{-1} \vert V^{(i)}} 
\end{equation}
which can be interpreted as $i$ being an unique isolated firm, albeit with an effective ``renormalized'' productivity $\tilde{z_i}=z_i - \braket{J_{i\leftarrow}\vert(\mathbf{M}^{(i)})^{-1}\vert J_{i\rightarrow}}<z_i$. The Hawkins-Simons condition implies that $\mathbf{M}^{(i)}$ must also fulfil the same conditions as $\mathbf{M}$, in particular $(\mathbf{M}^{(i)})^{-1}$ has positive components. For the whole economy to be functional, the effective productivity of each firm must be positive.

\subsection{Sectoral interpretation} 
\label{sub:sectoral_interpretation}
One can also extend this analysis to different economical sectors. Consider for simplicity two sectors, so that $\mathbf{M}$ has the following block structure:
\begin{equation}
    \mathbf{M} := \left(\begin{matrix}
    &\mathbf{M}_1& -\mathbf{J}_{12}& \\
    &-\mathbf{J}_{21}& \mathbf{M}_2 
    \end{matrix}\right)
\end{equation}
where the matrix $\mathbf{M}_1$ is made of firms from sector $1$ and interlinkages between them, while the $\mathbf{J}$ matrices links sector $1$ with sector $2$. Calling $\ket{P}=(\ket{P_1},\ket{P_2})$ and $\ket{V}=(\ket{V_1},\ket{V_2})$, we can write as in eq.\eqref{eq:schur_1firm} that
\begin{equation}
     \left(\mathbf{M}_1 - \mathbf{J}_{12}\mathbf{M}_2^{-1}\mathbf{J}_{21}\right)\ket{P_1} = \ket{V_1} + \mathbf{J}_{21} \mathbf{M}_2^{-1}\ket{V_2}
 \end{equation} 
and thus the Hawkins-Simon condition is fulfilled also if both $\mathbf{M}_2$ and $\mathbf{M}_1 - \mathbf{J}_{12}\mathbf{M}_2^{-1}\mathbf{J}_{21}$ are M-matrices for any choice of sector partitioning in the economy. In other words, it is not just that $\mathbf{M}_1$ must be an M-matrix by producing enough goods for the consumption of firms in sector $1$, but it must also produce enough goods for all of the other sectors.

\section{Real Input-Output Networks}
\label{sec:real_networks}
In this section we will attempt confront our model with available, but partial data. At this stage, this is more of an exercise that gives color to the general framework presented in the body of the paper.

Since our stability criterium depends on spectral properties of a matrix describing the entirety of the economy, we need in principle highly detailed network data to present a full analysis. Most available data, however, consists of input-output tables describing sale and purchase relationships between entities, be they firms or larger entities such as sectors or even countries, or simple relational data describing who is in a client-supplier relationship with whom, with the latter having a significantly larger coverage. While it allows to some degree to infer the importance of certain firms in the network, it does not correspond directly to the $J_{ij}$ coefficients that appear in our formalism. 

The total productivity factors $z_i$, which measure the efficiency with which a firm turns inputs into outputs, are also hard to deduce from actual production data. In particular, detailed output data is seldom available, and so productivity measures must be constructed from revenue data only, allowing for potential errors between the actual and the inferred productivity levels. For a detailed discussion,  see \cite{NBERw15712}.

Detailed data is becoming increasingly available, and we intend to show that spectral analysis of production networks is in principle possible. 

\subsection{Dataset and definitions} 
\label{sub:dataset}
We use the FactSet Supply Chain Relationships database to build a supply chain network. The FactSet dataset contains a list of relational data between firms, stating if firms $A$ and $B$ have a client/supplier relation, if they are in competition or if they have a joint venture. It is built by collecting information from primary public sources such as SEC 10-K annual filings, investor presentations and press releases, and covers about $23,000$ publicly traded companies with over $325,000$ relationships. Since the relationships are infered from data released to the public, we cannot be sure that it is an exhaustive database of all the relationships between firms, but the subset of relationships deemed important by the firm themselves. 

Such links between firms have a finite duration in time and have thus a beginning and end date. For our study, we have chosen the set of client/supplier relationships during the whole year of $2015$. This allows us to build a graph $\mathcal{G}$ where a link $i\to j$ exists whenever $i$ is reported to be a supplier of $j$ or when $j$ is reported to be a client of $i$. This graph $\mathcal{G}$ consists of $237$ weakly linked subgraphs \footnote{A weakly linked subgraph is such that any two nodes in it can be linked by a directed path.}, of which we select the largest subgraph $\mathcal{G}_0$. The graph $\mathcal{G}_0$ consists of $10,447$ firms with $40,300$ relationships between them. The remaining subgraphs are very small and consist of at most a few tens of firms. 

From $\mathcal{G}_0$ we construct an adjacency matrix $\mathbf{A}$ defined by
\begin{equation}\label{eq:adj_matrix_G0}
(\mathbf{A})_{ij}= \left\{ \begin{matrix}
1 &\mbox{ if } j\rightarrow i \in \mathcal{G}_0\\
0 &\mbox { otherwise}
\end{matrix}\right.
\end{equation}
which carries the topology we expect from the matrix $(\mathbf{M})_{ij}=z_i \delta_{ij}-J_{ij}$ defined in the main text. Indeed if for simplicity one supposes that all firms have the same productivity $z$ and that $J_{ij}=1$ for all links in the graph, then one has $\mathbf{M}=z\mathbf{1}-\mathbf{A}$. A full study of the properties of $\mathbf{M}$ would therefore require the supplementary knowledge of the values taken by the $J_{ij}$ and $z_i$ terms, i.e. have access to the dollar amount of products exchanged between firms and the total production of each firm. 

\subsection{Spectral study of the adjacency matrix} 
\label{sub:spectral_study_of_the_adjacency_matrix}
As the main body of our paper suggests, it is interesting to look into the properties of the eigenvalues and eigenvectors of the matrix $\mathbf{M}$, which under the hypothesis presented above are the same as those of the matrix $-\mathbf{A}$, up to a shift by $z$ on the real axis for the eigenvalues. 

A notion introduced in our paper is that of localized and delocalized eigenvectors. To quantify the localization properties we introduce the Inverse Participation Ratio, or IPR,\footnote{Akin to the Herfindahl index used in the economics litterature.} $H$ of a normalized complex-valued eigenvector $\ket{v^\lambda}=\left(v^\lambda_1,\ldots, v^\lambda_N\right)$ associated to an eigenvalue $\lambda$ as:
\begin{equation}
    H(\lambda) := \sum_{i} \left\vert\left\vert v_i^\lambda \right\vert\right\vert ^4 .
\end{equation}

Indeed, for a perfectly delocalized eigenvector we should have $\forall i,\quad v^\lambda_i=1/\sqrt{N}$ and therefore an $H(\lambda)=1/N$, while an eigenvector localized on a single site $\ket{v^\lambda}=(1,0,\ldots,0)$ has trivially a value $H(\lambda)=1$. It follows then that $L(\lambda)=1/H(\lambda)$ is a measure of the number of sites over which a given eigenvector associated to the eigenvalue $\lambda$ is spread out.

\begin{figure}[tb]
    \centering
    \includegraphics[width=18cm]{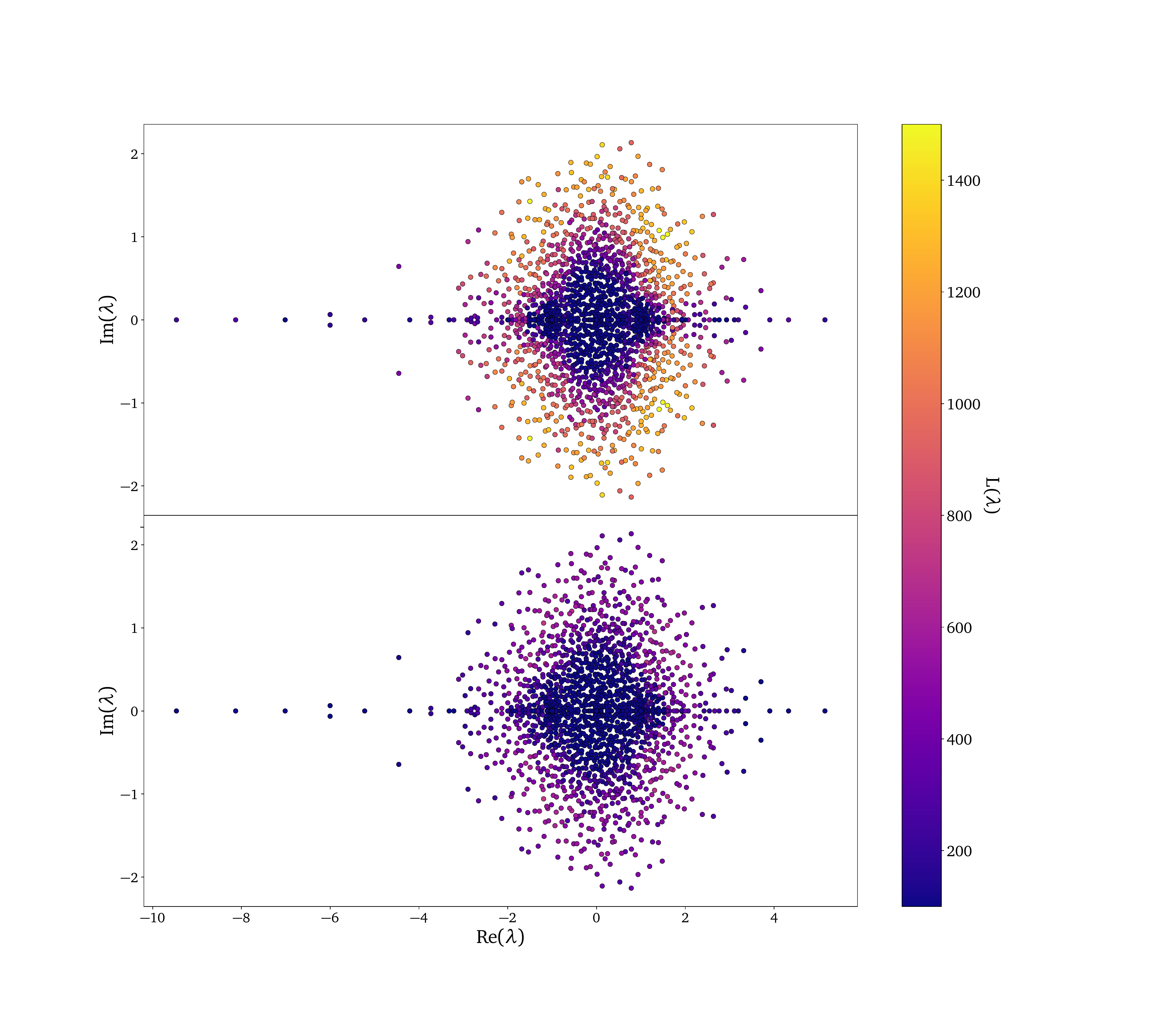}
    \caption{Spectrum of $-\mathbf{A}$ in the complex plane, with the colour of each eigenvalue $\lambda$ given by $L(\lambda)$, defined as the inverse of the Inverse Participation Ratio $H$ of the corresponding eigenvector, e.g. the number of firms over which the eigenvector is effectively spread. The color on the top plot corresponds to the right eigenvectors, while the one on the bottom corresponds to the left eigenvectors. Notice the different localization profiles on left and right eigenvectors.}
    \label{fig:factset_adj}
\end{figure}

We have computed the eigenvalues and eigenvectors of $-\mathbf{A}$, as visible on Figure~\ref{fig:factset_adj}, where one can see qualitatively that the spectral properties of the real adjacency matrix are not far from the simplistic random regular graph case shown in Figure~\ref{fig:ipr_smooth}: we see indeed a bulk to the right hand side with isolated eigenvectors to the left of it, and with varying localization properties of the eigenvectors accross the spectrum.

\begin{figure}[tb]
    \centering
    \includegraphics[trim=100 100 100 100, clip, width=12cm]{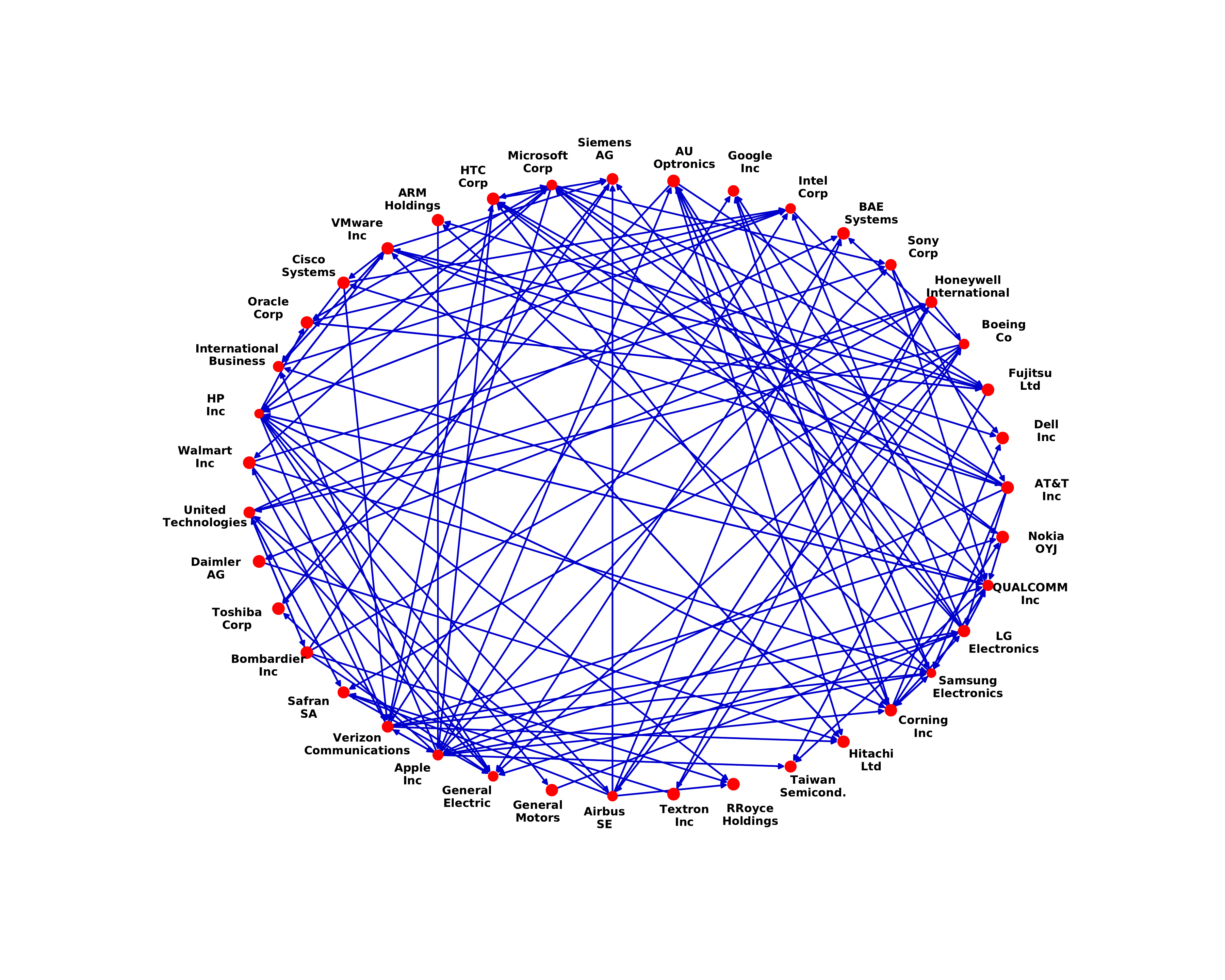}
    \caption{Subgraph of the $38$ firms with the largest contributions to $\ket{r^{\min}}$.}
    \label{fig:right_ev_graph}
\end{figure}

\begin{figure}[tb]
    \centering
    \includegraphics[trim=100 150 100 130, clip, width=12cm]{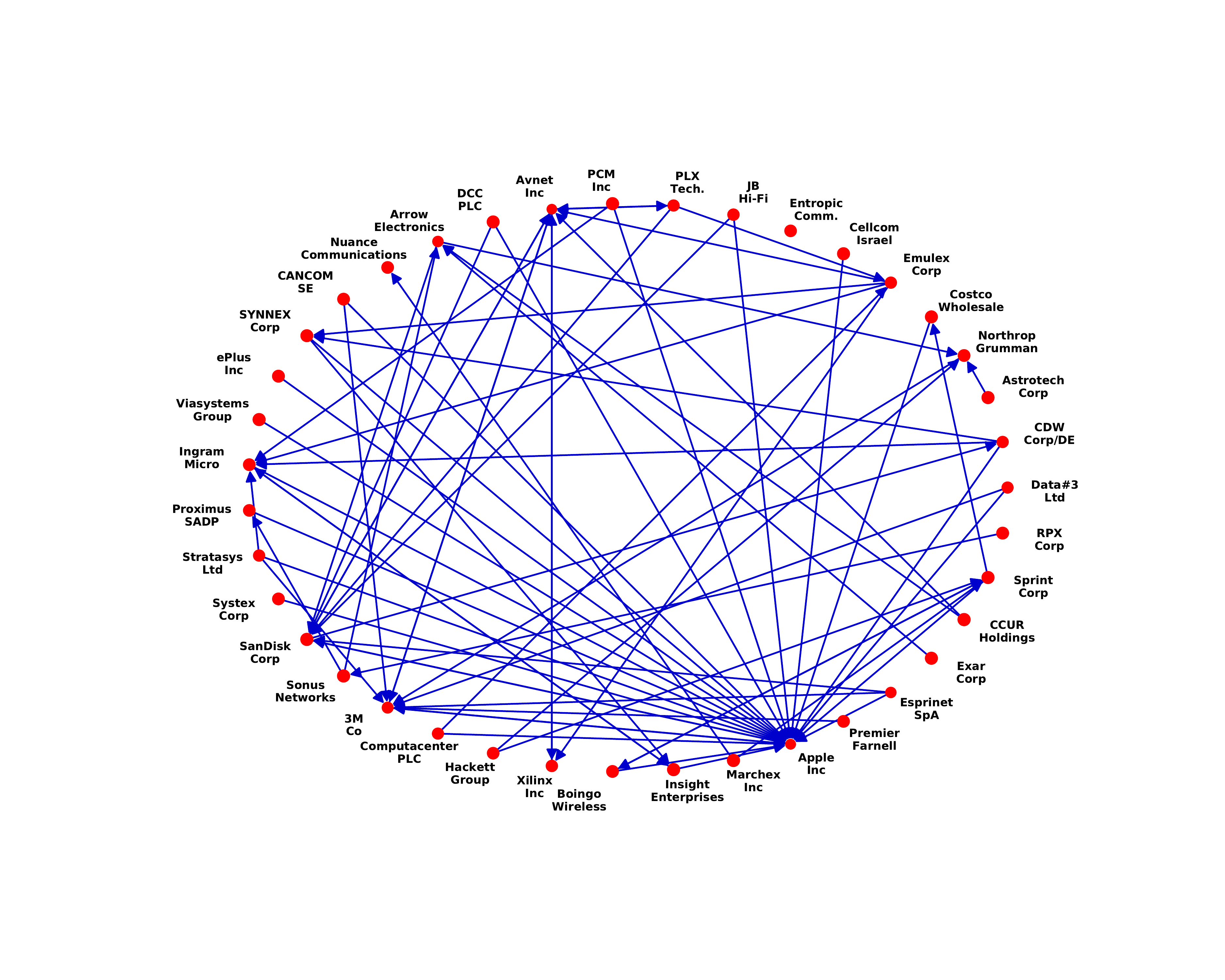}
    \caption{Subgraph of the $38$ firms with the largest contributions to $\ket{l^{\min}}$.}
    \label{fig:left_ev_graph}
\end{figure}

Of capital interest to our study are the left-most eigenvectors $\ket{l^{\min}}$ and $\ket{r^{\min}}$, which owing to the Perron-Frobenius theorem have real positive components and are associated to a real eigenvalue $\lambda_{\min}$, as can also be seen in Figure~\ref{fig:factset_adj}. In the graph we are studying, we have found that the eigenvectors are spread out over $L_r(\lambda_{\min})\simeq 40$ firms for the right eigenvector and $L_l(\lambda_{\min})\simeq 185$. For both eigenvectors, we have listed the $20$ firms with the most important contributions on Tables~\ref{tab:r_table} and~\ref{tab:l_table}, as well as their yearly reported sales for the year $2015$\footnote{In a few cases this data was not available.}. We have also represented the subgraph of the $38$ firms with the largest contributions to these eigenvectors and their interlinkages in Figures~\ref{fig:left_ev_graph} and~\ref{fig:right_ev_graph}. This allows for a better understanding of eq.~\eqref{eq:m_inv_eigen3}: if a shock hits firms represented in Figure~\ref{fig:right_ev_graph} this will be reflected in the prices of goods produced by firms represented in Figure~\ref{fig:left_ev_graph}. 
This corresponds fairly well to basic intuition, as firms in Figure~\ref{fig:right_ev_graph} correspond roughly to firms producing goods (such as electronics and software) that we expect to be purchased, after possible transformations along the supply chain, by retail and communication firms in Figure~\ref{fig:left_ev_graph}. The same holds by the inverting the roles of the firms, as idiosyncratic shocks to firms contributing to the left eigenvector will have an effect on the output of firms represented in the right eigenvector.

\begin{figure}[tb]
    \centering
    \includegraphics[width=10cm]{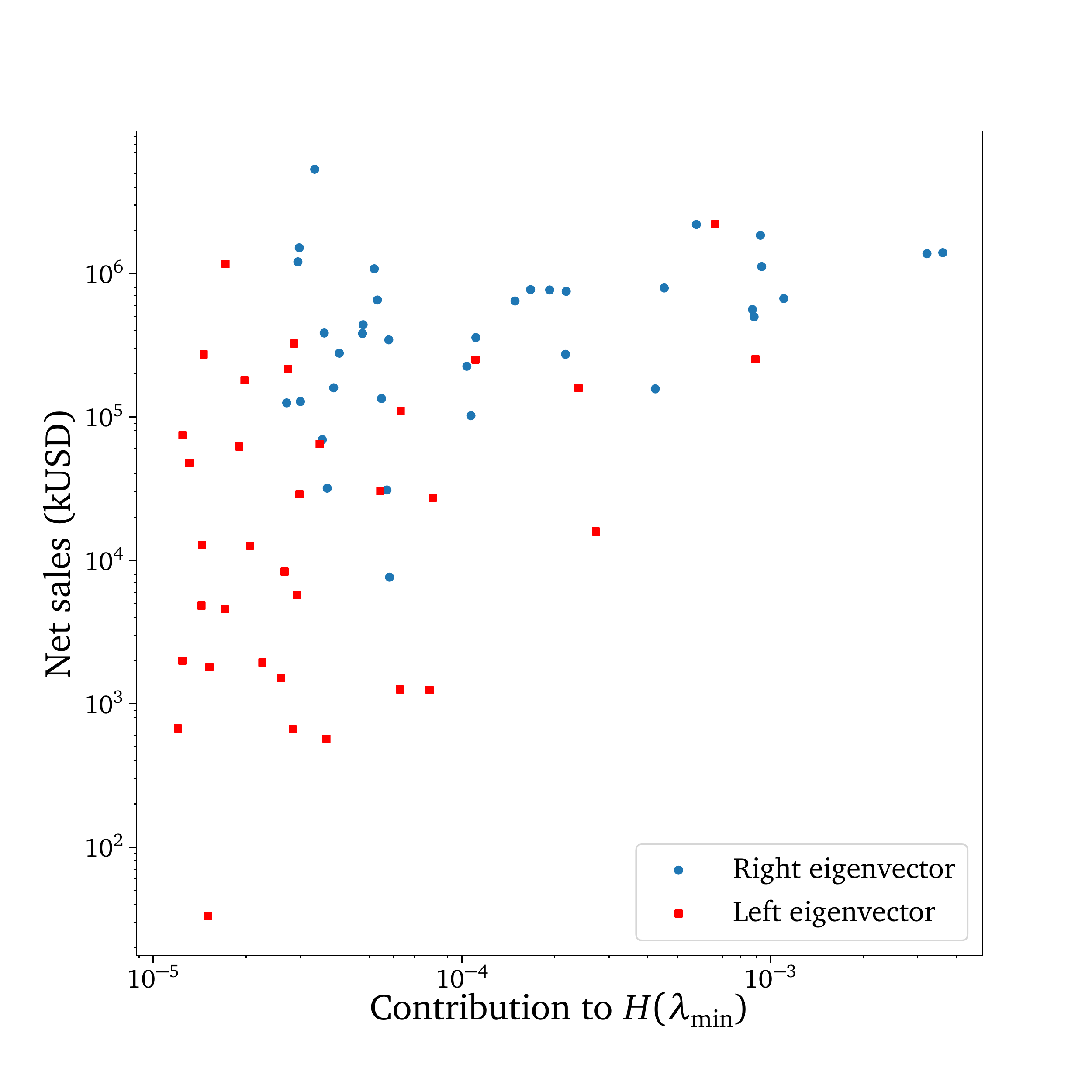}
    \caption{Log-log plot of the sales of the 38 main firms contributing to $\ket{v^{\lambda_{\min}}}$ vs. their contribution to $H(\lambda_{\min})$, suggesting that firms with a higher overlap with $\ket{v^{\lambda_{\min}}}$ have higher sales.}
    \label{fig:sales_v_ipr}
\end{figure}

Although the data available to us is not as complete as we could wish for a full study of the matrix $\mathbf{M}$, we think nevertheless that it supports some of our qualitative conclusions. For example, in the main text we argue that the power-law firm size distribution could be a feature explained by the proximity of an instability, with the largest firms being those that have a high overlap with the leftmost eigenvectors of $\mathbf{M}$. To test this, we plot the sales of the 36 firms with the largest contribution to $\ket{r^{\min}}$ and 34 firms with the largest contribution to $\ket{l^{\min}}$ against their contributions of these firms to the IPR/Herfindahl of these vectors\footnote{Which is none other than the quartic root of the scalar product between a firm's vector and $\ket{l,r^{\lambda_{\min}}}$.} in Figure~\ref{fig:sales_v_ipr}, suggesting an increasing relation between the overlap of a firm with $\ket{v^{\lambda_{\min}}}$ and its sales.

\begin{table}[]
\begin{center}
\begin{tabular}{|l|l|l|}
\hline
\textbf{Company Name}                     &  \textbf{Sales (kUSD)} & \textbf{Herfindahl Contribution} $\vert\vert r_i^{\lambda_{\min}}\vert\vert^4$ \\
\hline
                                   HP Inc & 1.40e+06 &          3.62e-03 \\
               Samsung Electronics Co Ltd & 1.37e+06 &          3.21e-03 \\
                                Airbus SE & 6.69e+05 &          1.10e-03 \\
                            Boeing Co/The & 1.12e+06 &          9.37e-04 \\
                      General Electric Co & 1.85e+06 &          9.28e-04 \\
                               Intel Corp & 4.99e+05 &          8.85e-04 \\
                           Microsoft Corp & 5.61e+05 &          8.74e-04 \\
                                Apple Inc & 2.20e+06 &          5.75e-04 \\
     International Business Machines Corp & 7.93e+05 &          4.53e-04 \\
                             QUALCOMM Inc & 1.57e+05 &          4.23e-04 \\
                               Google Inc & 7.51e+05 &          2.18e-04 \\
                                Sony Corp & 2.73e+05 &          2.17e-04 \\
                 United Technologies Corp & 7.68e+05 &          1.92e-04 \\
               Verizon Communications Inc & 7.73e+05 &          1.67e-04 \\
                               Siemens AG & 6.43e+05 &          1.49e-04 \\
              Honeywell International Inc & 3.58e+05 &          1.11e-04 \\
                                Safran SA & 1.02e+05 &          1.07e-04 \\
Taiwan Semiconductor Manufacturing Co Ltd & 2.25e+05 &          1.04e-04 \\
                         ARM Holdings PLC & 7.63e+03 &          5.83e-05 \\
                       LG Electronics Inc & 3.45e+05 &          5.80e-05 \\
\hline
\end{tabular}
\caption{Firms with the 20 most important contributions to the eigenvector $\ket{r^{\lambda_{\min}}}$, their respective contributions to the Herfindahl, and their reported sales for the year $2015$ in thousands of $2015$ USD. }
\label{tab:r_table}
\end{center}
\end{table}

\begin{table}[]
\begin{center}
\begin{tabular}{|l|l|l|}
\hline
\textbf{Company Name}                     &  \textbf{Sales (kUSD)} & \textbf{Herfindahl Contribution} $\vert\vert l_i^{\lambda_{\min}}\vert\vert^4$ \\
\hline
                   Avnet Inc & 2.52e+05 &          8.94e-04 \\
                   Apple Inc & 2.20e+06 &          6.61e-04 \\
                Esprinet SpA & 1.59e+04 &          2.72e-04 \\
       Arrow Electronics Inc & 1.59e+05 &          2.39e-04 \\
                       3M Co & 2.50e+05 &          1.11e-04 \\
           Computacenter PLC & 2.74e+04 &          8.06e-05 \\
                  Data\#3 Ltd & 1.25e+03 &          7.86e-05 \\
                 CDW Corp/DE & 1.11e+05 &          6.33e-05 \\
                 Emulex Corp & 1.26e+03 &          6.31e-05 \\
               Stratasys Ltd &      ? &          6.30e-05 \\
                  Xilinx Inc & 3.04e+04 &          5.45e-05 \\
                JB Hi-Fi Ltd &      ? &          5.19e-05 \\
 Entropic Communications Inc & 5.68e+02 &          3.64e-05 \\
               Proximus SADP & 6.49e+04 &          3.46e-05 \\
   Nuance Communications Inc & 2.89e+04 &          2.97e-05 \\
         Premier Farnell Ltd & 5.70e+03 &          2.92e-05 \\
            Ingram Micro Inc & 3.24e+05 &          2.86e-05 \\
                 Systex Corp & 6.62e+02 &          2.83e-05 \\
       Northrop Grumman Corp & 2.16e+05 &          2.73e-05 \\
                   CANCOM SE & 8.35e+03 &          2.67e-05 \\
\hline
\end{tabular}
\caption{Firms with the 20 most important contributions to the eigenvector $\ket{l^{\lambda_{\min}}}$, their respective contributions to the Herfindahl, and their reported sales for the year $2015$ in thousands of $2015$ USD. }
\label{tab:l_table}
\end{center}
\end{table}



\bibliography{biblio.bib}

 \newcommand{\noop}[1]{}
\begin{thebibliography}{71}%
\makeatletter
\providecommand \@ifxundefined [1]{%
 \@ifx{#1\undefined}
}%
\providecommand \@ifnum [1]{%
 \ifnum #1\expandafter \@firstoftwo
 \else \expandafter \@secondoftwo
 \fi
}%
\providecommand \@ifx [1]{%
 \ifx #1\expandafter \@firstoftwo
 \else \expandafter \@secondoftwo
 \fi
}%
\providecommand \natexlab [1]{#1}%
\providecommand \enquote  [1]{``#1''}%
\providecommand \bibnamefont  [1]{#1}%
\providecommand \bibfnamefont [1]{#1}%
\providecommand \citenamefont [1]{#1}%
\providecommand \href@noop [0]{\@secondoftwo}%
\providecommand \href [0]{\begingroup \@sanitize@url \@href}%
\providecommand \@href[1]{\@@startlink{#1}\@@href}%
\providecommand \@@href[1]{\endgroup#1\@@endlink}%
\providecommand \@sanitize@url [0]{\catcode `\\12\catcode `\$12\catcode
  `\&12\catcode `\#12\catcode `\^12\catcode `\_12\catcode `\%12\relax}%
\providecommand \@@startlink[1]{}%
\providecommand \@@endlink[0]{}%
\providecommand \url  [0]{\begingroup\@sanitize@url \@url }%
\providecommand \@url [1]{\endgroup\@href {#1}{\urlprefix }}%
\providecommand \urlprefix  [0]{URL }%
\providecommand \Eprint [0]{\href }%
\providecommand \doibase [0]{http://dx.doi.org/}%
\providecommand \selectlanguage [0]{\@gobble}%
\providecommand \bibinfo  [0]{\@secondoftwo}%
\providecommand \bibfield  [0]{\@secondoftwo}%
\providecommand \translation [1]{[#1]}%
\providecommand \BibitemOpen [0]{}%
\providecommand \bibitemStop [0]{}%
\providecommand \bibitemNoStop [0]{.\EOS\space}%
\providecommand \EOS [0]{\spacefactor3000\relax}%
\providecommand \BibitemShut  [1]{\csname bibitem#1\endcsname}%
\let\auto@bib@innerbib\@empty
\bibitem [{\citenamefont {Long}\ and\ \citenamefont
  {Plosser}(1983)}]{long1983real}%
  \BibitemOpen
  \bibfield  {author} {\bibinfo {author} {\bibfnamefont {John~B.}\ \bibnamefont
  {Long}}\ and\ \bibinfo {author} {\bibfnamefont {Charles~I.}\ \bibnamefont
  {Plosser}},\ }\bibfield  {title} {\enquote {\bibinfo {title} {Real business
  cycles},}\ }\href {\doibase 10.1086/261128} {\bibfield  {journal} {\bibinfo
  {journal} {Journal of Political Economy}\ }\textbf {\bibinfo {volume} {91}},\
  \bibinfo {pages} {39--69} (\bibinfo {year} {1983})}\BibitemShut {NoStop}%
\bibitem [{\citenamefont {Cochrane}(1994)}]{cochrane1994shocks}%
  \BibitemOpen
  \bibfield  {author} {\bibinfo {author} {\bibfnamefont {John}\ \bibnamefont
  {Cochrane}},\ }\href {\doibase 10.3386/w4698} {\emph {\bibinfo {title}
  {Shocks}}},\ \bibinfo {type} {Tech. Rep.}\ (\bibinfo {year}
  {1994})\BibitemShut {NoStop}%
\bibitem [{\citenamefont {Bernanke}\ \emph {et~al.}(1994)\citenamefont
  {Bernanke}, \citenamefont {Gertler},\ and\ \citenamefont
  {Gilchrist}}]{bernanke1994financial}%
  \BibitemOpen
  \bibfield  {author} {\bibinfo {author} {\bibfnamefont {Ben}\ \bibnamefont
  {Bernanke}}, \bibinfo {author} {\bibfnamefont {Mark}\ \bibnamefont
  {Gertler}}, \ and\ \bibinfo {author} {\bibfnamefont {Simon}\ \bibnamefont
  {Gilchrist}},\ }\href {\doibase 10.3386/w4789} {\emph {\bibinfo {title} {The
  Financial Accelerator and the Flight to Quality}}},\ \bibinfo {type} {Tech.
  Rep.}\ (\bibinfo {year} {1994})\BibitemShut {NoStop}%
\bibitem [{\citenamefont {Lucas}(1995)}]{Lucas}%
  \BibitemOpen
  \bibfield  {author} {\bibinfo {author} {\bibfnamefont {R.~E.}\ \bibnamefont
  {Lucas}},\ }\enquote {\bibinfo {title} {Understanding business cycles},}\ in\
  \href {\doibase 10.1007/978-1-349-24002-9_17} {\emph {\bibinfo {booktitle}
  {Essential Readings in Economics}}},\ \bibinfo {editor} {edited by\ \bibinfo
  {editor} {\bibfnamefont {Saul}\ \bibnamefont {Estrin}}\ and\ \bibinfo
  {editor} {\bibfnamefont {Alan}\ \bibnamefont {Marin}}}\ (\bibinfo
  {publisher} {Macmillan Education UK},\ \bibinfo {address} {London},\ \bibinfo
  {year} {1995})\ pp.\ \bibinfo {pages} {306--327}\BibitemShut {NoStop}%
\bibitem [{\citenamefont {Dupor}(1999)}]{dupor1999aggregation}%
  \BibitemOpen
  \bibfield  {author} {\bibinfo {author} {\bibfnamefont {Bill}\ \bibnamefont
  {Dupor}},\ }\bibfield  {title} {\enquote {\bibinfo {title} {Aggregation and
  irrelevance in multi-sector models},}\ }\href {\doibase
  10.1016/s0304-3932(98)00057-9} {\bibfield  {journal} {\bibinfo  {journal}
  {Journal of Monetary Economics}\ }\textbf {\bibinfo {volume} {43}},\ \bibinfo
  {pages} {391--409} (\bibinfo {year} {1999})}\BibitemShut {NoStop}%
\bibitem [{\citenamefont {Farmer}(1999)}]{farmer1999macroeconomics}%
  \BibitemOpen
  \bibfield  {author} {\bibinfo {author} {\bibfnamefont {R.E.A.}\ \bibnamefont
  {Farmer}},\ }\href {\doibase 10.1007/BF02299031} {\emph {\bibinfo {title}
  {The Macroeconomics of Self-fulfilling Prophecies}}},\ Macroeconomics of
  Self-fulfilling Prophecies\ (\bibinfo  {publisher} {MIT Press},\ \bibinfo
  {year} {1999})\BibitemShut {NoStop}%
\bibitem [{\citenamefont {Brock}\ and\ \citenamefont
  {Durlauf}(2001)}]{Brock_2001}%
  \BibitemOpen
  \bibfield  {author} {\bibinfo {author} {\bibfnamefont {W.~A.}\ \bibnamefont
  {Brock}}\ and\ \bibinfo {author} {\bibfnamefont {S.~N.}\ \bibnamefont
  {Durlauf}},\ }\bibfield  {title} {\enquote {\bibinfo {title} {Discrete choice
  with social interactions},}\ }\href {\doibase 10.1111/1467-937x.00168}
  {\bibfield  {journal} {\bibinfo  {journal} {The Review of Economic Studies}\
  }\textbf {\bibinfo {volume} {68}},\ \bibinfo {pages} {235--260} (\bibinfo
  {year} {2001})}\BibitemShut {NoStop}%
\bibitem [{\citenamefont {Bouchaud}(2013)}]{Bouchaud_2013}%
  \BibitemOpen
  \bibfield  {author} {\bibinfo {author} {\bibfnamefont {Jean-Philippe}\
  \bibnamefont {Bouchaud}},\ }\bibfield  {title} {\enquote {\bibinfo {title}
  {Crises and collective socio-economic phenomena: Simple models and
  challenges},}\ }\href {\doibase 10.1007/s10955-012-0687-3} {\bibfield
  {journal} {\bibinfo  {journal} {Journal of Statistical Physics}\ }\textbf
  {\bibinfo {volume} {151}},\ \bibinfo {pages} {567--606} (\bibinfo {year}
  {2013})}\BibitemShut {NoStop}%
\bibitem [{\citenamefont {Anand}\ \emph {et~al.}(2013)\citenamefont {Anand},
  \citenamefont {Kirman},\ and\ \citenamefont {Marsili}}]{Anand_2013}%
  \BibitemOpen
  \bibfield  {author} {\bibinfo {author} {\bibfnamefont {Kartik}\ \bibnamefont
  {Anand}}, \bibinfo {author} {\bibfnamefont {Alan}\ \bibnamefont {Kirman}}, \
  and\ \bibinfo {author} {\bibfnamefont {Matteo}\ \bibnamefont {Marsili}},\
  }\bibfield  {title} {\enquote {\bibinfo {title} {Epidemics of rules, rational
  negligence and market crashes},}\ }\href {\doibase
  10.1080/1351847x.2011.601872} {\bibfield  {journal} {\bibinfo  {journal} {The
  European Journal of Finance}\ }\textbf {\bibinfo {volume} {19}},\ \bibinfo
  {pages} {438--447} (\bibinfo {year} {2013})}\BibitemShut {NoStop}%
\bibitem [{\citenamefont {da~Gama~Batista}\ \emph {et~al.}(2015)\citenamefont
  {da~Gama~Batista}, \citenamefont {Bouchaud},\ and\ \citenamefont
  {Challet}}]{da_Gama_Batista_2015}%
  \BibitemOpen
  \bibfield  {author} {\bibinfo {author} {\bibfnamefont {Jo{\~{a}}o}\
  \bibnamefont {da~Gama~Batista}}, \bibinfo {author} {\bibfnamefont
  {Jean-Philippe}\ \bibnamefont {Bouchaud}}, \ and\ \bibinfo {author}
  {\bibfnamefont {Damien}\ \bibnamefont {Challet}},\ }\bibfield  {title}
  {\enquote {\bibinfo {title} {Sudden trust collapse in networked societies},}\
  }\href {\doibase 10.1140/epjb/e2015-50645-1} {\bibfield  {journal} {\bibinfo
  {journal} {The European Physical Journal B}\ }\textbf {\bibinfo {volume}
  {88}} (\bibinfo {year} {2015}),\ 10.1140/epjb/e2015-50645-1}\BibitemShut
  {NoStop}%
\bibitem [{\citenamefont {Gabaix}(2011)}]{gabaix2011granular}%
  \BibitemOpen
  \bibfield  {author} {\bibinfo {author} {\bibfnamefont {Xavier}\ \bibnamefont
  {Gabaix}},\ }\bibfield  {title} {\enquote {\bibinfo {title} {The granular
  origins of aggregate fluctuations},}\ }\href {\doibase 10.3982/ecta8769}
  {\bibfield  {journal} {\bibinfo  {journal} {Econometrica}\ }\textbf {\bibinfo
  {volume} {79}},\ \bibinfo {pages} {733--772} (\bibinfo {year}
  {2011})}\BibitemShut {NoStop}%
\bibitem [{\citenamefont {Wyart}\ and\ \citenamefont
  {Bouchaud}(2003)}]{wyart2003statistical}%
  \BibitemOpen
  \bibfield  {author} {\bibinfo {author} {\bibfnamefont {Matthieu}\
  \bibnamefont {Wyart}}\ and\ \bibinfo {author} {\bibfnamefont {Jean-Philippe}\
  \bibnamefont {Bouchaud}},\ }\bibfield  {title} {\enquote {\bibinfo {title}
  {Statistical models for company growth},}\ }\href {\doibase
  10.1016/s0378-4371(03)00267-x} {\bibfield  {journal} {\bibinfo  {journal}
  {Physica A: Statistical Mechanics and its Applications}\ }\textbf {\bibinfo
  {volume} {326}},\ \bibinfo {pages} {241--255} (\bibinfo {year}
  {2003})}\BibitemShut {NoStop}%
\bibitem [{\citenamefont {Sutton}(2002)}]{sutton2002variance}%
  \BibitemOpen
  \bibfield  {author} {\bibinfo {author} {\bibfnamefont {John}\ \bibnamefont
  {Sutton}},\ }\bibfield  {title} {\enquote {\bibinfo {title} {The variance of
  firm growth rates: the `scaling' puzzle},}\ }\href {\doibase
  10.1016/s0378-4371(02)00852-x} {\bibfield  {journal} {\bibinfo  {journal}
  {Physica A: Statistical Mechanics and its Applications}\ }\textbf {\bibinfo
  {volume} {312}},\ \bibinfo {pages} {577--590} (\bibinfo {year}
  {2002})}\BibitemShut {NoStop}%
\bibitem [{\citenamefont {di~Giovanni}\ \emph {et~al.}(2017)\citenamefont
  {di~Giovanni}, \citenamefont {Levchenko},\ and\ \citenamefont
  {Mejean}}]{di_Giovanni_2017}%
  \BibitemOpen
  \bibfield  {author} {\bibinfo {author} {\bibfnamefont {Julian}\ \bibnamefont
  {di~Giovanni}}, \bibinfo {author} {\bibfnamefont {Andrei~A.}\ \bibnamefont
  {Levchenko}}, \ and\ \bibinfo {author} {\bibfnamefont {Isabelle}\
  \bibnamefont {Mejean}},\ }\bibfield  {title} {\enquote {\bibinfo {title}
  {Large firms and international business cycle comovement},}\ }\href {\doibase
  10.1257/aer.p20171006} {\bibfield  {journal} {\bibinfo  {journal} {American
  Economic Review}\ }\textbf {\bibinfo {volume} {107}},\ \bibinfo {pages}
  {598--602} (\bibinfo {year} {2017})}\BibitemShut {NoStop}%
\bibitem [{\citenamefont {Acemoglu}\ \emph {et~al.}(2012)\citenamefont
  {Acemoglu}, \citenamefont {Carvalho}, \citenamefont {Ozdaglar},\ and\
  \citenamefont {Tahbaz-Salehi}}]{acemoglu2012network}%
  \BibitemOpen
  \bibfield  {author} {\bibinfo {author} {\bibfnamefont {Daron}\ \bibnamefont
  {Acemoglu}}, \bibinfo {author} {\bibfnamefont {Vasco}\ \bibnamefont
  {Carvalho}}, \bibinfo {author} {\bibfnamefont {Asu}\ \bibnamefont
  {Ozdaglar}}, \ and\ \bibinfo {author} {\bibfnamefont {Alireza}\ \bibnamefont
  {Tahbaz-Salehi}},\ }\bibfield  {title} {\enquote {\bibinfo {title} {The
  network origins of aggregate fluctuations},}\ }\href {\doibase
  10.3982/ecta9623} {\bibfield  {journal} {\bibinfo  {journal} {Econometrica}\
  }\textbf {\bibinfo {volume} {80}},\ \bibinfo {pages} {1977--2016} (\bibinfo
  {year} {2012})}\BibitemShut {NoStop}%
\bibitem [{\citenamefont {Acemoglu}\ \emph {et~al.}(2016)\citenamefont
  {Acemoglu}, \citenamefont {Ozdaglar},\ and\ \citenamefont
  {Tahbaz-Salehi}}]{acemoglu2015networks}%
  \BibitemOpen
  \bibfield  {author} {\bibinfo {author} {\bibfnamefont {Daron}\ \bibnamefont
  {Acemoglu}}, \bibinfo {author} {\bibfnamefont {Asu}\ \bibnamefont
  {Ozdaglar}}, \ and\ \bibinfo {author} {\bibfnamefont {Alireza}\ \bibnamefont
  {Tahbaz-Salehi}},\ }\href {\doibase 10.1093/oxfordhb/9780199948277.013.17}
  {\emph {\bibinfo {title} {Networks, Shocks, and Systemic Risk}}},\ edited by\
  \bibinfo {editor} {\bibfnamefont {Yann}\ \bibnamefont {Bramoull{\'{e}}}},
  \bibinfo {editor} {\bibfnamefont {Andrea}\ \bibnamefont {Galeotti}}, \ and\
  \bibinfo {editor} {\bibfnamefont {Brian}\ \bibnamefont {Rogers}}\ (\bibinfo
  {publisher} {Oxford University Press},\ \bibinfo {year} {2016})\BibitemShut
  {NoStop}%
\bibitem [{\citenamefont {Acemoglu}\ \emph {et~al.}(2015)\citenamefont
  {Acemoglu}, \citenamefont {Akcigit},\ and\ \citenamefont
  {Kerr}}]{acemoglu2016networks}%
  \BibitemOpen
  \bibfield  {author} {\bibinfo {author} {\bibfnamefont {Daron}\ \bibnamefont
  {Acemoglu}}, \bibinfo {author} {\bibfnamefont {Ufuk}\ \bibnamefont
  {Akcigit}}, \ and\ \bibinfo {author} {\bibfnamefont {William~R.}\
  \bibnamefont {Kerr}},\ }\bibfield  {title} {\enquote {\bibinfo {title}
  {Networks and the macroeconomy: An empirical exploration},}\ }\href {\doibase
  10.2139/ssrn.2630102} {\bibfield  {journal} {\bibinfo  {journal} {{SSRN}
  Electronic Journal}\ } (\bibinfo {year} {2015}),\
  10.2139/ssrn.2630102}\BibitemShut {NoStop}%
\bibitem [{\citenamefont {Haldane}\ and\ \citenamefont
  {May}(2011)}]{HaldaneMay2011}%
  \BibitemOpen
  \bibfield  {author} {\bibinfo {author} {\bibfnamefont {Andrew~G.}\
  \bibnamefont {Haldane}}\ and\ \bibinfo {author} {\bibfnamefont {Robert~M.}\
  \bibnamefont {May}},\ }\bibfield  {title} {\enquote {\bibinfo {title}
  {Systemic risk in banking ecosystems},}\ }\href {\doibase
  10.1038/nature09659} {\bibfield  {journal} {\bibinfo  {journal} {Nature}\
  }\textbf {\bibinfo {volume} {469}},\ \bibinfo {pages} {351--355} (\bibinfo
  {year} {2011})}\BibitemShut {NoStop}%
\bibitem [{\citenamefont {Gai}\ and\ \citenamefont {Kapadia}(2010)}]{Gai2010}%
  \BibitemOpen
  \bibfield  {author} {\bibinfo {author} {\bibfnamefont {Prasanna}\
  \bibnamefont {Gai}}\ and\ \bibinfo {author} {\bibfnamefont {Sujit}\
  \bibnamefont {Kapadia}},\ }\bibfield  {title} {\enquote {\bibinfo {title}
  {Contagion in financial networks},}\ }\href {\doibase 10.2139/ssrn.1577043}
  {\bibfield  {journal} {\bibinfo  {journal} {{SSRN} Electronic Journal}\ }
  (\bibinfo {year} {2010}),\ 10.2139/ssrn.1577043}\BibitemShut {NoStop}%
\bibitem [{\citenamefont {Tasca}\ and\ \citenamefont
  {Battiston}(2011)}]{Tasca_2011}%
  \BibitemOpen
  \bibfield  {author} {\bibinfo {author} {\bibfnamefont {Paolo}\ \bibnamefont
  {Tasca}}\ and\ \bibinfo {author} {\bibfnamefont {Stefano}\ \bibnamefont
  {Battiston}},\ }\bibfield  {title} {\enquote {\bibinfo {title}
  {Diversification and financial stability},}\ }\href {\doibase
  10.2139/ssrn.1878596} {\bibfield  {journal} {\bibinfo  {journal} {{SSRN}
  Electronic Journal}\ } (\bibinfo {year} {2011}),\
  10.2139/ssrn.1878596}\BibitemShut {NoStop}%
\bibitem [{\citenamefont {Caccioli}\ \emph {et~al.}(2017)\citenamefont
  {Caccioli}, \citenamefont {Barucca},\ and\ \citenamefont
  {Kobayashi}}]{Caccioli2017}%
  \BibitemOpen
  \bibfield  {author} {\bibinfo {author} {\bibfnamefont {Fabio}\ \bibnamefont
  {Caccioli}}, \bibinfo {author} {\bibfnamefont {Paolo}\ \bibnamefont
  {Barucca}}, \ and\ \bibinfo {author} {\bibfnamefont {Teruyoshi}\ \bibnamefont
  {Kobayashi}},\ }\bibfield  {title} {\enquote {\bibinfo {title} {Network
  models of financial systemic risk: A review},}\ }\href {\doibase
  10.2139/ssrn.3066722} {\bibfield  {journal} {\bibinfo  {journal} {{SSRN}
  Electronic Journal}\ } (\bibinfo {year} {2017}),\
  10.2139/ssrn.3066722}\BibitemShut {NoStop}%
\bibitem [{\citenamefont {Foerster}\ \emph {et~al.}(2008)\citenamefont
  {Foerster}, \citenamefont {Sarte},\ and\ \citenamefont
  {Watson}}]{Foerster_2008}%
  \BibitemOpen
  \bibfield  {author} {\bibinfo {author} {\bibfnamefont {Andrew~T.}\
  \bibnamefont {Foerster}}, \bibinfo {author} {\bibfnamefont
  {Pierre-Daniel~G.}\ \bibnamefont {Sarte}}, \ and\ \bibinfo {author}
  {\bibfnamefont {Mark~W.}\ \bibnamefont {Watson}},\ }\bibfield  {title}
  {\enquote {\bibinfo {title} {Sectoral vs. aggregate shocks: A structural
  factor analysis of industrial production},}\ }\href {\doibase
  10.2139/ssrn.2187904} {\bibfield  {journal} {\bibinfo  {journal} {{SSRN}
  Electronic Journal}\ } (\bibinfo {year} {2008}),\
  10.2139/ssrn.2187904}\BibitemShut {NoStop}%
\bibitem [{\citenamefont {Carvalho}\ and\ \citenamefont
  {Gabaix}(2010)}]{Carvalho_2010}%
  \BibitemOpen
  \bibfield  {author} {\bibinfo {author} {\bibfnamefont {Vasco~M.}\
  \bibnamefont {Carvalho}}\ and\ \bibinfo {author} {\bibfnamefont {Xavier}\
  \bibnamefont {Gabaix}},\ }\bibfield  {title} {\enquote {\bibinfo {title} {The
  great diversification and its undoing},}\ }\href {\doibase
  10.2139/ssrn.1684196} {\bibfield  {journal} {\bibinfo  {journal} {{SSRN}
  Electronic Journal}\ } (\bibinfo {year} {2010}),\
  10.2139/ssrn.1684196}\BibitemShut {NoStop}%
\bibitem [{\citenamefont {Acemoglu}\ \emph {et~al.}(2013)\citenamefont
  {Acemoglu}, \citenamefont {Ozdaglar},\ and\ \citenamefont
  {Tahbaz-Salehi}}]{acemoglu2013network}%
  \BibitemOpen
  \bibfield  {author} {\bibinfo {author} {\bibfnamefont {Daron}\ \bibnamefont
  {Acemoglu}}, \bibinfo {author} {\bibfnamefont {Asuman}\ \bibnamefont
  {Ozdaglar}}, \ and\ \bibinfo {author} {\bibfnamefont {Alireza}\ \bibnamefont
  {Tahbaz-Salehi}},\ }\href {\doibase 10.3386/w19230} {\emph {\bibinfo {title}
  {The Network Origins of Large Economic Downturns}}},\ \bibinfo {type} {Tech.
  Rep.}\ (\bibinfo {year} {2013})\BibitemShut {NoStop}%
\bibitem [{\citenamefont {Bonart}\ \emph {et~al.}(2014)\citenamefont {Bonart},
  \citenamefont {Bouchaud}, \citenamefont {Landier},\ and\ \citenamefont
  {Thesmar}}]{bonart2014instabilities}%
  \BibitemOpen
  \bibfield  {author} {\bibinfo {author} {\bibfnamefont {Julius}\ \bibnamefont
  {Bonart}}, \bibinfo {author} {\bibfnamefont {Jean-Philippe}\ \bibnamefont
  {Bouchaud}}, \bibinfo {author} {\bibfnamefont {Augustin}\ \bibnamefont
  {Landier}}, \ and\ \bibinfo {author} {\bibfnamefont {David}\ \bibnamefont
  {Thesmar}},\ }\bibfield  {title} {\enquote {\bibinfo {title} {Instabilities
  in large economies: aggregate volatility without idiosyncratic shocks},}\
  }\href {\doibase 10.1088/1742-5468/2014/10/p10040} {\bibfield  {journal}
  {\bibinfo  {journal} {Journal of Statistical Mechanics: Theory and
  Experiment}\ }\textbf {\bibinfo {volume} {2014}},\ \bibinfo {pages} {P10040}
  (\bibinfo {year} {2014})}\BibitemShut {NoStop}%
\bibitem [{\citenamefont {Hawkins}(1948)}]{hawkins1948some}%
  \BibitemOpen
  \bibfield  {author} {\bibinfo {author} {\bibfnamefont {David}\ \bibnamefont
  {Hawkins}},\ }\bibfield  {title} {\enquote {\bibinfo {title} {Some conditions
  of macroeconomic stability},}\ }\href {\doibase 10.2307/1909272} {\bibfield
  {journal} {\bibinfo  {journal} {Econometrica}\ }\textbf {\bibinfo {volume}
  {16}},\ \bibinfo {pages} {309} (\bibinfo {year} {1948})}\BibitemShut
  {NoStop}%
\bibitem [{\citenamefont {Hawkins}\ and\ \citenamefont
  {Simon}(1949)}]{hawkins1949note}%
  \BibitemOpen
  \bibfield  {author} {\bibinfo {author} {\bibfnamefont {David}\ \bibnamefont
  {Hawkins}}\ and\ \bibinfo {author} {\bibfnamefont {Herbert~A.}\ \bibnamefont
  {Simon}},\ }\bibfield  {title} {\enquote {\bibinfo {title} {Note: Some
  conditions of macroeconomic stability},}\ }\href {\doibase 10.2307/1905526}
  {\bibfield  {journal} {\bibinfo  {journal} {Econometrica}\ }\textbf {\bibinfo
  {volume} {17}},\ \bibinfo {pages} {245} (\bibinfo {year} {1949})}\BibitemShut
  {NoStop}%
\bibitem [{\citenamefont {Bak}\ \emph {et~al.}(1992)\citenamefont {Bak},
  \citenamefont {Chen}, \citenamefont {Scheinkman},\ and\ \citenamefont
  {Woodford}}]{bak1993aggregate}%
  \BibitemOpen
  \bibfield  {author} {\bibinfo {author} {\bibfnamefont {Per}\ \bibnamefont
  {Bak}}, \bibinfo {author} {\bibfnamefont {Kan}\ \bibnamefont {Chen}},
  \bibinfo {author} {\bibfnamefont {Jos\'{e}}\ \bibnamefont {Scheinkman}}, \
  and\ \bibinfo {author} {\bibfnamefont {Michael}\ \bibnamefont {Woodford}},\
  }\bibfield  {title} {\enquote {\bibinfo {title} {Aggregate fluctuations from
  independent sectoral shocks: Self-organized criticality in a model of
  production and inventory dynamics},}\ }\href {\doibase 10.3386/w4241} {\
  (\bibinfo {year} {1992}),\ 10.3386/w4241}\BibitemShut {NoStop}%
\bibitem [{\citenamefont {Scheinkman}\ and\ \citenamefont
  {Woodford}(1994)}]{scheinkman1994self}%
  \BibitemOpen
  \bibfield  {author} {\bibinfo {author} {\bibfnamefont {Jos\'{e}}\
  \bibnamefont {Scheinkman}}\ and\ \bibinfo {author} {\bibfnamefont {Michael}\
  \bibnamefont {Woodford}},\ }\bibfield  {title} {\enquote {\bibinfo {title}
  {Self-organized criticality and economic fluctuations},}\ }\href
  {https://EconPapers.repec.org/RePEc:aea:aecrev:v:84:y:1994:i:2:p:417-21}
  {\bibfield  {journal} {\bibinfo  {journal} {American Economic Review}\
  }\textbf {\bibinfo {volume} {84}},\ \bibinfo {pages} {417--21} (\bibinfo
  {year} {1994})}\BibitemShut {NoStop}%
\bibitem [{\citenamefont {Bak}(2013)}]{bak2013nature}%
  \BibitemOpen
  \bibfield  {author} {\bibinfo {author} {\bibfnamefont {Per}\ \bibnamefont
  {Bak}},\ }\href {\doibase 10.1007/978-1-4757-5426-1} {\emph {\bibinfo {title}
  {How nature works: the science of self-organized criticality}}}\ (\bibinfo
  {publisher} {Springer Science \& Business Media},\ \bibinfo {year}
  {2013})\BibitemShut {NoStop}%
\bibitem [{\citenamefont {May}(1972)}]{may1972will}%
  \BibitemOpen
  \bibfield  {author} {\bibinfo {author} {\bibfnamefont {Robert~M.}\
  \bibnamefont {May}},\ }\bibfield  {title} {\enquote {\bibinfo {title} {Will a
  large complex system be stable?}}\ }\href {\doibase 10.1038/238413a0}
  {\bibfield  {journal} {\bibinfo  {journal} {Nature}\ }\textbf {\bibinfo
  {volume} {238}},\ \bibinfo {pages} {413--414} (\bibinfo {year}
  {1972})}\BibitemShut {NoStop}%
\bibitem [{\citenamefont {Bunin}(2017)}]{Bunin_2017}%
  \BibitemOpen
  \bibfield  {author} {\bibinfo {author} {\bibfnamefont {Guy}\ \bibnamefont
  {Bunin}},\ }\bibfield  {title} {\enquote {\bibinfo {title} {Ecological
  communities with lotka-volterra dynamics},}\ }\href {\doibase
  10.1103/PhysRevE.95.042414} {\bibfield  {journal} {\bibinfo  {journal} {Phys.
  Rev. E}\ }\textbf {\bibinfo {volume} {95}},\ \bibinfo {pages} {042414}
  (\bibinfo {year} {2017})}\BibitemShut {NoStop}%
\bibitem [{\citenamefont {Bunin}(2016)}]{bunin2016interaction}%
  \BibitemOpen
  \bibfield  {author} {\bibinfo {author} {\bibfnamefont {Guy}\ \bibnamefont
  {Bunin}},\ }\bibfield  {title} {\enquote {\bibinfo {title} {Interaction
  patterns and diversity in assembled ecological communities},}\ }\href@noop {}
  {\  (\bibinfo {year} {2016})},\ \Eprint
  {http://arxiv.org/abs/arXiv:1607.04734} {arXiv:1607.04734} \BibitemShut
  {NoStop}%
\bibitem [{\citenamefont {Biroli}\ \emph {et~al.}(2018)\citenamefont {Biroli},
  \citenamefont {Bunin},\ and\ \citenamefont
  {Cammarota}}]{biroli2018marginally}%
  \BibitemOpen
  \bibfield  {author} {\bibinfo {author} {\bibfnamefont {Giulio}\ \bibnamefont
  {Biroli}}, \bibinfo {author} {\bibfnamefont {Guy}\ \bibnamefont {Bunin}}, \
  and\ \bibinfo {author} {\bibfnamefont {Chiara}\ \bibnamefont {Cammarota}},\
  }\bibfield  {title} {\enquote {\bibinfo {title} {Marginally stable equilibria
  in critical ecosystems},}\ }\href {\doibase 10.1088/1367-2630/aada58}
  {\bibfield  {journal} {\bibinfo  {journal} {New Journal of Physics}\ }\textbf
  {\bibinfo {volume} {20}},\ \bibinfo {pages} {083051} (\bibinfo {year}
  {2018})}\BibitemShut {NoStop}%
\bibitem [{\citenamefont {Nirei}\ and\ \citenamefont
  {Scheinkman}(2019)}]{Nirei_2019}%
  \BibitemOpen
  \bibfield  {author} {\bibinfo {author} {\bibfnamefont {Makoto}\ \bibnamefont
  {Nirei}}\ and\ \bibinfo {author} {\bibfnamefont {Jos{\'{e}}}\ \bibnamefont
  {Scheinkman}},\ }\bibfield  {title} {\enquote {\bibinfo {title}
  {Self-organization of inflation volatility},}\ }\href {\doibase
  10.2139/ssrn.3396528} {\bibfield  {journal} {\bibinfo  {journal} {{SSRN}
  Electronic Journal}\ } (\bibinfo {year} {2019}),\
  10.2139/ssrn.3396528}\BibitemShut {NoStop}%
\bibitem [{\citenamefont {Moran}\ and\ \citenamefont {Bouchaud}()}]{ustocome}%
  \BibitemOpen
  \bibfield  {author} {\bibinfo {author} {\bibfnamefont {Jos\'{e}}\
  \bibnamefont {Moran}}\ and\ \bibinfo {author} {\bibfnamefont {Jean-Philippe}\
  \bibnamefont {Bouchaud}},\ }\href@noop {} {}\bibinfo {note} {In
  preparation}\BibitemShut {NoStop}%
\bibitem [{\citenamefont {Raval}(2019)}]{Raval_2019}%
  \BibitemOpen
  \bibfield  {author} {\bibinfo {author} {\bibfnamefont {Devesh~R.}\
  \bibnamefont {Raval}},\ }\bibfield  {title} {\enquote {\bibinfo {title} {The
  micro elasticity of substitution and non-neutral technology},}\ }\href
  {\doibase 10.1111/1756-2171.12265} {\bibfield  {journal} {\bibinfo  {journal}
  {The {RAND} Journal of Economics}\ } (\bibinfo {year} {2019}),\
  10.1111/1756-2171.12265}\BibitemShut {NoStop}%
\bibitem [{\citenamefont {Reed}\ and\ \citenamefont
  {Simon}(2011)}]{reed_simon_2011}%
  \BibitemOpen
  \bibfield  {author} {\bibinfo {author} {\bibfnamefont {John}\ \bibnamefont
  {Reed}}\ and\ \bibinfo {author} {\bibfnamefont {Bernard}\ \bibnamefont
  {Simon}},\ }\href
  {https://www.ft.com/content/9ac4d7e2-595f-11e0-bc39-00144feab49a} {\enquote
  {\bibinfo {title} {Car components hit by {J}apan aftershock},}\ } (\bibinfo
  {year} {2011}),\ \bibinfo {note} {checked: 24.01.2019}\BibitemShut {NoStop}%
\bibitem [{\citenamefont {Tajitsu}(2016)}]{reuters_japan_2011}%
  \BibitemOpen
  \bibfield  {author} {\bibinfo {author} {\bibfnamefont {Naomi}\ \bibnamefont
  {Tajitsu}},\ }\href
  {https://www.reuters.com/article/us-japan-quake-supplychain-idUSKCN0WW09N}
  {\enquote {\bibinfo {title} {Five years after {J}apan quake, rewiring of auto
  supply chain hits limits},}\ } (\bibinfo {year} {2016}),\ \bibinfo {note}
  {checked: 24.01.2019}\BibitemShut {NoStop}%
\bibitem [{\citenamefont {Fisher}(2011)}]{fisher_hbs_2011}%
  \BibitemOpen
  \bibfield  {author} {\bibinfo {author} {\bibfnamefont {Dennis}\ \bibnamefont
  {Fisher}},\ }\href
  {https://hbswk.hbs.edu/item/japan-disaster-shakes-up-supply-chain-strategies}
  {\enquote {\bibinfo {title} {{J}apan disaster shakes up supply-chain
  strategies},}\ } (\bibinfo {year} {2011}),\ \bibinfo {note} {checked:
  24.01.2019}\BibitemShut {NoStop}%
\bibitem [{\citenamefont {Colon}\ and\ \citenamefont {Bouchaud}()}]{colon2019}%
  \BibitemOpen
  \bibfield  {author} {\bibinfo {author} {\bibfnamefont {C\'{e}lian}\
  \bibnamefont {Colon}}\ and\ \bibinfo {author} {\bibfnamefont {Jean-Philippe}\
  \bibnamefont {Bouchaud}},\ }\href@noop {} {\enquote {\bibinfo {title}
  {Transition from plasticity to instability in the structure of economic
  networks},}\ }\bibinfo {note} {In preparation}\BibitemShut {NoStop}%
\bibitem [{\citenamefont {Fiedler}\ and\ \citenamefont
  {Ptak}(1962)}]{fiedler1962matrices}%
  \BibitemOpen
  \bibfield  {author} {\bibinfo {author} {\bibfnamefont {Miroslav}\
  \bibnamefont {Fiedler}}\ and\ \bibinfo {author} {\bibfnamefont {Vlastimil}\
  \bibnamefont {Ptak}},\ }\bibfield  {title} {\enquote {\bibinfo {title} {On
  matrices with non-positive off-diagonal elements and positive principal
  minors},}\ }\href {http://eudml.org/doc/12135} {\bibfield  {journal}
  {\bibinfo  {journal} {Czechoslovak Mathematical Journal}\ }\textbf {\bibinfo
  {volume} {12}},\ \bibinfo {pages} {382--400} (\bibinfo {year}
  {1962})}\BibitemShut {NoStop}%
\bibitem [{\citenamefont {Girko}(1985)}]{Girko_1985}%
  \BibitemOpen
  \bibfield  {author} {\bibinfo {author} {\bibfnamefont {V.~L.}\ \bibnamefont
  {Girko}},\ }\bibfield  {title} {\enquote {\bibinfo {title} {Circular law},}\
  }\href {\doibase 10.1137/1129095} {\bibfield  {journal} {\bibinfo  {journal}
  {Theory of Probability {\&} Its Applications}\ }\textbf {\bibinfo {volume}
  {29}},\ \bibinfo {pages} {694--706} (\bibinfo {year} {1985})}\BibitemShut
  {NoStop}%
\bibitem [{\citenamefont {Tao}\ and\ \citenamefont {Vu}(2008)}]{Tao_2008}%
  \BibitemOpen
  \bibfield  {author} {\bibinfo {author} {\bibfnamefont {Terence}\ \bibnamefont
  {Tao}}\ and\ \bibinfo {author} {\bibfnamefont {Van}\ \bibnamefont {Vu}},\
  }\bibfield  {title} {\enquote {\bibinfo {title} {Random matrices: the
  circular law},}\ }\href {\doibase 10.1142/s0219199708002788} {\bibfield
  {journal} {\bibinfo  {journal} {Communications in Contemporary Mathematics}\
  }\textbf {\bibinfo {volume} {10}},\ \bibinfo {pages} {261--307} (\bibinfo
  {year} {2008})}\BibitemShut {NoStop}%
\bibitem [{\citenamefont {Metz}\ \emph {et~al.}(2018)\citenamefont {Metz},
  \citenamefont {Neri},\ and\ \citenamefont {Rogers}}]{metz2018spectra}%
  \BibitemOpen
  \bibfield  {author} {\bibinfo {author} {\bibfnamefont {Fernando~Lucas}\
  \bibnamefont {Metz}}, \bibinfo {author} {\bibfnamefont {Izaak}\ \bibnamefont
  {Neri}}, \ and\ \bibinfo {author} {\bibfnamefont {Tim}\ \bibnamefont
  {Rogers}},\ }\bibfield  {title} {\enquote {\bibinfo {title} {Spectra of
  sparse non-hermitian random matrices},}\ }\href@noop {} {\  (\bibinfo {year}
  {2018})},\ \Eprint {http://arxiv.org/abs/arXiv:1811.10416} {arXiv:1811.10416}
  \BibitemShut {NoStop}%
\bibitem [{\citenamefont {Wigner}(1967)}]{wigner1967random}%
  \BibitemOpen
  \bibfield  {author} {\bibinfo {author} {\bibfnamefont {Eugene~P.}\
  \bibnamefont {Wigner}},\ }\bibfield  {title} {\enquote {\bibinfo {title}
  {Random matrices in physics},}\ }\href {\doibase 10.1137/1009001} {\bibfield
  {journal} {\bibinfo  {journal} {{SIAM} Review}\ }\textbf {\bibinfo {volume}
  {9}},\ \bibinfo {pages} {1--23} (\bibinfo {year} {1967})}\BibitemShut
  {NoStop}%
\bibitem [{\citenamefont {Biroli}\ and\ \citenamefont
  {Monasson}(1999)}]{biroli1999single}%
  \BibitemOpen
  \bibfield  {author} {\bibinfo {author} {\bibfnamefont {G}~\bibnamefont
  {Biroli}}\ and\ \bibinfo {author} {\bibfnamefont {R}~\bibnamefont
  {Monasson}},\ }\bibfield  {title} {\enquote {\bibinfo {title} {A single
  defect approximation for localized states on random lattices},}\ }\href
  {\doibase 10.1088/0305-4470/32/24/101} {\bibfield  {journal} {\bibinfo
  {journal} {Journal of Physics A: Mathematical and General}\ }\textbf
  {\bibinfo {volume} {32}},\ \bibinfo {pages} {L255--L261} (\bibinfo {year}
  {1999})}\BibitemShut {NoStop}%
\bibitem [{\citenamefont {Farkas}\ \emph {et~al.}(2011)\citenamefont {Farkas},
  \citenamefont {Derenyi}, \citenamefont {Barab{\'a}si},\ and\ \citenamefont
  {Vicsek}}]{farkas2001spectra}%
  \BibitemOpen
  \bibfield  {author} {\bibinfo {author} {\bibfnamefont {Illes~J.}\
  \bibnamefont {Farkas}}, \bibinfo {author} {\bibfnamefont {Imre}\ \bibnamefont
  {Derenyi}}, \bibinfo {author} {\bibfnamefont {A.-L.}\ \bibnamefont
  {Barab{\'a}si}}, \ and\ \bibinfo {author} {\bibfnamefont {Tamas}\
  \bibnamefont {Vicsek}},\ }\bibfield  {title} {\enquote {\bibinfo {title}
  {Spectra of "real-world" graphs: Beyond the semicircle law},}\ }\bibfield
  {booktitle} {\emph {\bibinfo {booktitle} {The Structure and Dynamics of
  Networks}},\ }\href {\doibase 10.1515/9781400841356.372} {\  (\bibinfo {year}
  {2011}),\ 10.1515/9781400841356.372}\BibitemShut {NoStop}%
\bibitem [{\citenamefont {Albert}\ and\ \citenamefont
  {Barab{\'{a}}si}(2002)}]{albert2002statistical}%
  \BibitemOpen
  \bibfield  {author} {\bibinfo {author} {\bibfnamefont {R{\'{e}}ka}\
  \bibnamefont {Albert}}\ and\ \bibinfo {author} {\bibfnamefont
  {Albert-L{\'{a}}szl{\'{o}}}\ \bibnamefont {Barab{\'{a}}si}},\ }\bibfield
  {title} {\enquote {\bibinfo {title} {Statistical mechanics of complex
  networks},}\ }\href {\doibase 10.1103/revmodphys.74.47} {\bibfield  {journal}
  {\bibinfo  {journal} {Reviews of Modern Physics}\ }\textbf {\bibinfo {volume}
  {74}},\ \bibinfo {pages} {47--97} (\bibinfo {year} {2002})}\BibitemShut
  {NoStop}%
\bibitem [{\citenamefont {Rogers}\ and\ \citenamefont
  {Castillo}(2009)}]{rogers2009cavity}%
  \BibitemOpen
  \bibfield  {author} {\bibinfo {author} {\bibfnamefont {Tim}\ \bibnamefont
  {Rogers}}\ and\ \bibinfo {author} {\bibfnamefont {Isaac~P\'erez}\
  \bibnamefont {Castillo}},\ }\bibfield  {title} {\enquote {\bibinfo {title}
  {Cavity approach to the spectral density of non-hermitian sparse matrices},}\
  }\href {\doibase 10.1103/PhysRevE.79.012101} {\bibfield  {journal} {\bibinfo
  {journal} {Phys. Rev. E}\ }\textbf {\bibinfo {volume} {79}},\ \bibinfo
  {pages} {012101} (\bibinfo {year} {2009})}\BibitemShut {NoStop}%
\bibitem [{\citenamefont {Fyodorov}\ and\ \citenamefont
  {Mirlin}(1991)}]{fyodorov1991localization}%
  \BibitemOpen
  \bibfield  {author} {\bibinfo {author} {\bibfnamefont {Yan}\ \bibnamefont
  {Fyodorov}}\ and\ \bibinfo {author} {\bibfnamefont {Alexander}\ \bibnamefont
  {Mirlin}},\ }\bibfield  {title} {\enquote {\bibinfo {title} {Localization in
  ensemble of sparse random matrices},}\ }\href {\doibase
  10.1103/physrevlett.67.2049} {\bibfield  {journal} {\bibinfo  {journal}
  {Physical Review Letters}\ }\textbf {\bibinfo {volume} {67}},\ \bibinfo
  {pages} {2049--2052} (\bibinfo {year} {1991})}\BibitemShut {NoStop}%
\bibitem [{\citenamefont {K\"{u}hn}(2008)}]{Kuhn_2008}%
  \BibitemOpen
  \bibfield  {author} {\bibinfo {author} {\bibfnamefont {Reimer}\ \bibnamefont
  {K\"{u}hn}},\ }\bibfield  {title} {\enquote {\bibinfo {title} {Spectra of
  sparse random matrices},}\ }\href {\doibase 10.1088/1751-8113/41/29/295002}
  {\bibfield  {journal} {\bibinfo  {journal} {Journal of Physics A:
  Mathematical and Theoretical}\ }\textbf {\bibinfo {volume} {41}},\ \bibinfo
  {pages} {295002} (\bibinfo {year} {2008})}\BibitemShut {NoStop}%
\bibitem [{\citenamefont {Benaych-Georges}\ and\ \citenamefont
  {P{\'{e}}ch{\'{e}}}(2014)}]{Benaych_Georges_2014}%
  \BibitemOpen
  \bibfield  {author} {\bibinfo {author} {\bibfnamefont {Florent}\ \bibnamefont
  {Benaych-Georges}}\ and\ \bibinfo {author} {\bibfnamefont {Sandrine}\
  \bibnamefont {P{\'{e}}ch{\'{e}}}},\ }\bibfield  {title} {\enquote {\bibinfo
  {title} {Largest eigenvalues and eigenvectors of band or sparse random
  matrices},}\ }\href {\doibase 10.1214/ecp.v19-3027} {\bibfield  {journal}
  {\bibinfo  {journal} {Electronic Communications in Probability}\ }\textbf
  {\bibinfo {volume} {19}} (\bibinfo {year} {2014}),\
  10.1214/ecp.v19-3027}\BibitemShut {NoStop}%
\bibitem [{\citenamefont {Neri}\ and\ \citenamefont
  {Metz}(2016)}]{neri2016eigenvalue}%
  \BibitemOpen
  \bibfield  {author} {\bibinfo {author} {\bibfnamefont {Izaak}\ \bibnamefont
  {Neri}}\ and\ \bibinfo {author} {\bibfnamefont {Fernando~Lucas}\ \bibnamefont
  {Metz}},\ }\bibfield  {title} {\enquote {\bibinfo {title} {Eigenvalue
  outliers of non-hermitian random matrices with a local tree structure},}\
  }\href {\doibase 10.1103/PhysRevLett.117.224101} {\bibfield  {journal}
  {\bibinfo  {journal} {Phys. Rev. Lett.}\ }\textbf {\bibinfo {volume} {117}},\
  \bibinfo {pages} {224101} (\bibinfo {year} {2016})}\BibitemShut {NoStop}%
\bibitem [{\citenamefont {Biroli}\ \emph {et~al.}(2010)\citenamefont {Biroli},
  \citenamefont {Semerjian},\ and\ \citenamefont
  {Tarzia}}]{biroli2010anderson}%
  \BibitemOpen
  \bibfield  {author} {\bibinfo {author} {\bibfnamefont {Giulio}\ \bibnamefont
  {Biroli}}, \bibinfo {author} {\bibfnamefont {Guilhem}\ \bibnamefont
  {Semerjian}}, \ and\ \bibinfo {author} {\bibfnamefont {Marco}\ \bibnamefont
  {Tarzia}},\ }\bibfield  {title} {\enquote {\bibinfo {title} {Anderson model
  on bethe lattices: Density of states, localization properties and isolated
  eigenvalue},}\ }\href {\doibase 10.1143/ptps.184.187} {\bibfield  {journal}
  {\bibinfo  {journal} {Progress of Theoretical Physics Supplement}\ }\textbf
  {\bibinfo {volume} {184}},\ \bibinfo {pages} {187--199} (\bibinfo {year}
  {2010})}\BibitemShut {NoStop}%
\bibitem [{\citenamefont {Chakrabarti}\ \emph {et~al.}(2008)\citenamefont
  {Chakrabarti}, \citenamefont {Wang}, \citenamefont {Wang}, \citenamefont
  {Leskovec},\ and\ \citenamefont {Faloutsos}}]{Chakrabarti_2008}%
  \BibitemOpen
  \bibfield  {author} {\bibinfo {author} {\bibfnamefont {Deepayan}\
  \bibnamefont {Chakrabarti}}, \bibinfo {author} {\bibfnamefont {Yang}\
  \bibnamefont {Wang}}, \bibinfo {author} {\bibfnamefont {Chenxi}\ \bibnamefont
  {Wang}}, \bibinfo {author} {\bibfnamefont {Jurij}\ \bibnamefont {Leskovec}},
  \ and\ \bibinfo {author} {\bibfnamefont {Christos}\ \bibnamefont
  {Faloutsos}},\ }\bibfield  {title} {\enquote {\bibinfo {title} {Epidemic
  thresholds in real networks},}\ }\href {\doibase 10.1145/1284680.1284681}
  {\bibfield  {journal} {\bibinfo  {journal} {{ACM} Transactions on Information
  and System Security}\ }\textbf {\bibinfo {volume} {10}},\ \bibinfo {pages}
  {1--26} (\bibinfo {year} {2008})}\BibitemShut {NoStop}%
\bibitem [{\citenamefont {Newman}(2010)}]{Newman_2010}%
  \BibitemOpen
  \bibfield  {author} {\bibinfo {author} {\bibfnamefont {Mark}\ \bibnamefont
  {Newman}},\ }\href@noop {} {\emph {\bibinfo {title} {Networks: An
  Introduction}}}\ (\bibinfo  {publisher} {Oxford University Press, Inc.},\
  \bibinfo {address} {New York, NY, USA},\ \bibinfo {year} {2010})\BibitemShut
  {NoStop}%
\bibitem [{\citenamefont {Castellano}\ and\ \citenamefont
  {Pastor-Satorras}(2017)}]{Castellano_2017}%
  \BibitemOpen
  \bibfield  {author} {\bibinfo {author} {\bibfnamefont {Claudio}\ \bibnamefont
  {Castellano}}\ and\ \bibinfo {author} {\bibfnamefont {Romualdo}\ \bibnamefont
  {Pastor-Satorras}},\ }\bibfield  {title} {\enquote {\bibinfo {title}
  {Relating topological determinants of complex networks to their spectral
  properties: Structural and dynamical effects},}\ }\href {\doibase
  10.1103/PhysRevX.7.041024} {\bibfield  {journal} {\bibinfo  {journal} {Phys.
  Rev. X}\ }\textbf {\bibinfo {volume} {7}},\ \bibinfo {pages} {041024}
  (\bibinfo {year} {2017})}\BibitemShut {NoStop}%
\bibitem [{\citenamefont {Baik}\ \emph {et~al.}(2005)\citenamefont {Baik},
  \citenamefont {Arous},\ and\ \citenamefont
  {P{\'{e}}ch{\'{e}}}}]{baik2005phase}%
  \BibitemOpen
  \bibfield  {author} {\bibinfo {author} {\bibfnamefont {Jinho}\ \bibnamefont
  {Baik}}, \bibinfo {author} {\bibfnamefont {G{\'{e}}rard~Ben}\ \bibnamefont
  {Arous}}, \ and\ \bibinfo {author} {\bibfnamefont {Sandrine}\ \bibnamefont
  {P{\'{e}}ch{\'{e}}}},\ }\bibfield  {title} {\enquote {\bibinfo {title} {Phase
  transition of the largest eigenvalue for nonnull complex sample covariance
  matrices},}\ }\href {\doibase 10.1214/009117905000000233} {\bibfield
  {journal} {\bibinfo  {journal} {The Annals of Probability}\ }\textbf
  {\bibinfo {volume} {33}},\ \bibinfo {pages} {1643--1697} (\bibinfo {year}
  {2005})}\BibitemShut {NoStop}%
\bibitem [{\citenamefont {Lifshitz}(1964)}]{Lifshitz_1964}%
  \BibitemOpen
  \bibfield  {author} {\bibinfo {author} {\bibfnamefont {I.M.}\ \bibnamefont
  {Lifshitz}},\ }\bibfield  {title} {\enquote {\bibinfo {title} {The energy
  spectrum of disordered systems},}\ }\href {\doibase
  10.1080/00018736400101061} {\bibfield  {journal} {\bibinfo  {journal}
  {Advances in Physics}\ }\textbf {\bibinfo {volume} {13}},\ \bibinfo {pages}
  {483--536} (\bibinfo {year} {1964})}\BibitemShut {NoStop}%
\bibitem [{\citenamefont {Thouless}(1974)}]{Thouless_1974}%
  \BibitemOpen
  \bibfield  {author} {\bibinfo {author} {\bibfnamefont {D.J.}\ \bibnamefont
  {Thouless}},\ }\bibfield  {title} {\enquote {\bibinfo {title} {Electrons in
  disordered systems and the theory of localization},}\ }\href {\doibase
  10.1016/0370-1573(74)90029-5} {\bibfield  {journal} {\bibinfo  {journal}
  {Physics Reports}\ }\textbf {\bibinfo {volume} {13}},\ \bibinfo {pages}
  {93--142} (\bibinfo {year} {1974})}\BibitemShut {NoStop}%
\bibitem [{\citenamefont {Sethna}(2006)}]{sethna2006statistical}%
  \BibitemOpen
  \bibfield  {author} {\bibinfo {author} {\bibfnamefont {J.}~\bibnamefont
  {Sethna}},\ }\href@noop {} {\emph {\bibinfo {title} {Statistical Mechanics:
  Entropy, Order Parameters and Complexity}}},\ Oxford Master Series in
  Physics\ (\bibinfo  {publisher} {OUP Oxford},\ \bibinfo {year}
  {2006})\BibitemShut {NoStop}%
\bibitem [{\citenamefont {Axtell}(2001)}]{Axtell_2001}%
  \BibitemOpen
  \bibfield  {author} {\bibinfo {author} {\bibfnamefont {R.~L.}\ \bibnamefont
  {Axtell}},\ }\bibfield  {title} {\enquote {\bibinfo {title} {Zipf
  distribution of u.s. firm sizes},}\ }\href {\doibase 10.1126/science.1062081}
  {\bibfield  {journal} {\bibinfo  {journal} {Science}\ }\textbf {\bibinfo
  {volume} {293}},\ \bibinfo {pages} {1818--1820} (\bibinfo {year}
  {2001})}\BibitemShut {NoStop}%
\bibitem [{\citenamefont {Doussal}\ \emph {et~al.}(2010)\citenamefont
  {Doussal}, \citenamefont {M\"{u}ller},\ and\ \citenamefont
  {Wiese}}]{le2010avalanches}%
  \BibitemOpen
  \bibfield  {author} {\bibinfo {author} {\bibfnamefont {P.~Le}\ \bibnamefont
  {Doussal}}, \bibinfo {author} {\bibfnamefont {M.}~\bibnamefont {M\"{u}ller}},
  \ and\ \bibinfo {author} {\bibfnamefont {K.~J.}\ \bibnamefont {Wiese}},\
  }\bibfield  {title} {\enquote {\bibinfo {title} {Avalanches in mean-field
  models and the barkhausen noise in spin-glasses},}\ }\href {\doibase
  10.1209/0295-5075/91/57004} {\bibfield  {journal} {\bibinfo  {journal} {{EPL}
  (Europhysics Letters)}\ }\textbf {\bibinfo {volume} {91}},\ \bibinfo {pages}
  {57004} (\bibinfo {year} {2010})}\BibitemShut {NoStop}%
\bibitem [{\citenamefont {Charbonneau}\ \emph {et~al.}(2014)\citenamefont
  {Charbonneau}, \citenamefont {Kurchan}, \citenamefont {Parisi}, \citenamefont
  {Urbani},\ and\ \citenamefont {Zamponi}}]{charbonneau2014fractal}%
  \BibitemOpen
  \bibfield  {author} {\bibinfo {author} {\bibfnamefont {Patrick}\ \bibnamefont
  {Charbonneau}}, \bibinfo {author} {\bibfnamefont {Jorge}\ \bibnamefont
  {Kurchan}}, \bibinfo {author} {\bibfnamefont {Giorgio}\ \bibnamefont
  {Parisi}}, \bibinfo {author} {\bibfnamefont {Pierfrancesco}\ \bibnamefont
  {Urbani}}, \ and\ \bibinfo {author} {\bibfnamefont {Francesco}\ \bibnamefont
  {Zamponi}},\ }\bibfield  {title} {\enquote {\bibinfo {title} {Fractal free
  energy landscapes in structural glasses},}\ }\href {\doibase
  10.1038/ncomms4725} {\bibfield  {journal} {\bibinfo  {journal} {Nature
  Communications}\ }\textbf {\bibinfo {volume} {5}} (\bibinfo {year} {2014}),\
  10.1038/ncomms4725}\BibitemShut {NoStop}%
\bibitem [{\citenamefont {M\"{u}ller}\ and\ \citenamefont
  {Wyart}(2015)}]{muller2015marginal}%
  \BibitemOpen
  \bibfield  {author} {\bibinfo {author} {\bibfnamefont {Markus}\ \bibnamefont
  {M\"{u}ller}}\ and\ \bibinfo {author} {\bibfnamefont {Matthieu}\ \bibnamefont
  {Wyart}},\ }\bibfield  {title} {\enquote {\bibinfo {title} {Marginal
  stability in structural, spin, and electron glasses},}\ }\href {\doibase
  10.1146/annurev-conmatphys-031214-014614} {\bibfield  {journal} {\bibinfo
  {journal} {Annual Review of Condensed Matter Physics}\ }\textbf {\bibinfo
  {volume} {6}},\ \bibinfo {pages} {177--200} (\bibinfo {year}
  {2015})}\BibitemShut {NoStop}%
\bibitem [{\citenamefont {Atalay}\ \emph {et~al.}(2011)\citenamefont {Atalay},
  \citenamefont {Horta{\c c}su}, \citenamefont {Roberts},\ and\ \citenamefont
  {Syverson}}]{atalay2011}%
  \BibitemOpen
  \bibfield  {author} {\bibinfo {author} {\bibfnamefont {Enghin}\ \bibnamefont
  {Atalay}}, \bibinfo {author} {\bibfnamefont {Ali}\ \bibnamefont {Horta{\c
  c}su}}, \bibinfo {author} {\bibfnamefont {James}\ \bibnamefont {Roberts}}, \
  and\ \bibinfo {author} {\bibfnamefont {Chad}\ \bibnamefont {Syverson}},\
  }\bibfield  {title} {\enquote {\bibinfo {title} {Network structure of
  production},}\ }\href {\doibase 10.1073/pnas.1015564108} {\bibfield
  {journal} {\bibinfo  {journal} {Proceedings of the National Academy of
  Sciences}\ } (\bibinfo {year} {2011}),\ 10.1073/pnas.1015564108},\ \Eprint
  {http://arxiv.org/abs/https://www.pnas.org/content/early/2011/03/07/1015564108.full.pdf}
  {https://www.pnas.org/content/early/2011/03/07/1015564108.full.pdf}
  \BibitemShut {NoStop}%
\bibitem [{\citenamefont {Perotti}\ \emph {et~al.}(2009)\citenamefont
  {Perotti}, \citenamefont {Billoni}, \citenamefont {Tamarit}, \citenamefont
  {Chialvo},\ and\ \citenamefont {Cannas}}]{Perotti2009}%
  \BibitemOpen
  \bibfield  {author} {\bibinfo {author} {\bibfnamefont {Juan~I.}\ \bibnamefont
  {Perotti}}, \bibinfo {author} {\bibfnamefont {Orlando~V.}\ \bibnamefont
  {Billoni}}, \bibinfo {author} {\bibfnamefont {Francisco~A.}\ \bibnamefont
  {Tamarit}}, \bibinfo {author} {\bibfnamefont {Dante~R.}\ \bibnamefont
  {Chialvo}}, \ and\ \bibinfo {author} {\bibfnamefont {Sergio~A.}\ \bibnamefont
  {Cannas}},\ }\bibfield  {title} {\enquote {\bibinfo {title} {Emergent
  self-organized complex network topology out of stability constraints},}\
  }\href {\doibase 10.1103/PhysRevLett.103.108701} {\bibfield  {journal}
  {\bibinfo  {journal} {Phys. Rev. Lett.}\ }\textbf {\bibinfo {volume} {103}},\
  \bibinfo {pages} {108701} (\bibinfo {year} {2009})}\BibitemShut {NoStop}%
\bibitem [{\citenamefont {De~Martino}\ \emph {et~al.}(2007)\citenamefont
  {De~Martino}, \citenamefont {Marsili},\ and\ \citenamefont {{P\'{e}rez
  Castillo}}}]{demartino_2007}%
  \BibitemOpen
  \bibfield  {author} {\bibinfo {author} {\bibfnamefont {Andrea}\ \bibnamefont
  {De~Martino}}, \bibinfo {author} {\bibfnamefont {Matteo}\ \bibnamefont
  {Marsili}}, \ and\ \bibinfo {author} {\bibfnamefont {Isaac}\ \bibnamefont
  {{P\'{e}rez Castillo}}},\ }\bibfield  {title} {\enquote {\bibinfo {title}
  {Typical properties of large random economies with linear activities},}\
  }\href {\doibase 10.1017/S1365100507060191} {\bibfield  {journal} {\bibinfo
  {journal} {Macroeconomic Dynamics}\ }\textbf {\bibinfo {volume} {11}},\
  \bibinfo {pages} {34–61} (\bibinfo {year} {2007})}\BibitemShut {NoStop}%
\bibitem [{\citenamefont {Bardoscia}\ \emph {et~al.}(2017)\citenamefont
  {Bardoscia}, \citenamefont {Livan},\ and\ \citenamefont
  {Marsili}}]{Bardoscia_2017}%
  \BibitemOpen
  \bibfield  {author} {\bibinfo {author} {\bibfnamefont {Marco}\ \bibnamefont
  {Bardoscia}}, \bibinfo {author} {\bibfnamefont {Giacomo}\ \bibnamefont
  {Livan}}, \ and\ \bibinfo {author} {\bibfnamefont {Matteo}\ \bibnamefont
  {Marsili}},\ }\bibfield  {title} {\enquote {\bibinfo {title} {Statistical
  mechanics of complex economies},}\ }\href {\doibase 10.1088/1742-5468/aa6688}
  {\bibfield  {journal} {\bibinfo  {journal} {Journal of Statistical Mechanics:
  Theory and Experiment}\ }\textbf {\bibinfo {volume} {2017}},\ \bibinfo
  {pages} {043401} (\bibinfo {year} {2017})}\BibitemShut {NoStop}%
\bibitem [{\citenamefont {Syverson}(2010)}]{NBERw15712}%
  \BibitemOpen
  \bibfield  {author} {\bibinfo {author} {\bibfnamefont {Chad}\ \bibnamefont
  {Syverson}},\ }\href {\doibase 10.3386/w15712} {\emph {\bibinfo {title} {What
  Determines Productivity?}}},\ \bibinfo {type} {Working Paper}\ \bibinfo
  {number} {15712}\ (\bibinfo  {institution} {National Bureau of Economic
  Research},\ \bibinfo {year} {2010})\BibitemShut {NoStop}%
\end{thebibliography}%

\end{document}